\documentclass[twocolumn]{aastex63}

\received{February 1, 2021}
\revised{March 1, 2021}
\accepted{\today}
\submitjournal{AJ}

\shorttitle{Gas and stellar kinematics in the Rosette Nebula}
\shortauthors{Lim et al.}
\begin{document}

\title{A kinematic perspective on the formation process of the 
stellar groups in the Rosette Nebula}

\correspondingauthor{Beomdu Lim}
\email{blim@khu.ac.kr}

\author[0000-0001-5797-9828]{Beomdu Lim}
\affiliation{School of Space Research, Kyung Hee University 1732, Deogyeong-daero, Giheung-gu, Yongin-si, Gyeonggi-do 17104, Republic of Korea}
\affiliation{Korea Astronomy and Space Science Institute, 776 Daedeokdae-ro, Yuseong-gu, Daejeon 34055, Republic of Korea}

\author[0000-0003-4071-9346]{Ya\"el Naz\'e}
\thanks{FNRS Senior Research Associate}
\affiliation{Space sciences, Technologies and Astrophysics Research Institute, 
Universit\'e de Li\`ege, Quartier Agora, All\'ee du 6 Ao\^ut 19c, B\^at. B5c, 4000, Li\`ege, Belgium} 

\author[0000-0002-5097-8707]{Jongsuk Hong}
\affiliation{Korea Astronomy and Space Science Institute, 776 Daedeokdae-ro, Yuseong-gu, Daejeon 34055, Republic of Korea}

\author[0000-0002-6982-7722]{Byeong-Gon Park}
\affiliation{Korea Astronomy and Space Science Institute, 776 Daedeokdae-ro, Yuseong-gu, Daejeon 34055, Republic of Korea}
\author[0000-0001-6842-1555]{Hyeong-Sik Yun}
\affiliation{School of Space Research, Kyung Hee University 1732, Deogyeong-daero, Giheung-gu, Yongin-si, Gyeonggi-do 17104, Republic of Korea}
\author[0000-0003-0537-5461]{Hee-Weon Yi}
\affiliation{School of Space Research, Kyung Hee University 1732, Deogyeong-daero, Giheung-gu, Yongin-si, Gyeonggi-do 17104, Republic of Korea}
\author[0000-0003-4099-1171]{Sunkyung Park}
\affiliation{Konkoly Observatory, Research Centre for Astronomy and Earth Sciences, E\"otv\"os Lor\'and Research Network (ELKH), Konkoly-Thege Mikl\'os \'ut 15-17, 1121 Budapest, Hungary}

\author[0000-0002-2013-1273]{Narae Hwang}
\affiliation{Korea Astronomy and Space Science Institute, 776 Daedeokdae-ro, Yuseong-gu, Daejeon 34055, Republic of Korea}

\author[0000-0003-3119-2087]{Jeong-Eun Lee}
\affiliation{School of Space Research, Kyung Hee University 1732, Deogyeong-daero, Giheung-gu, Yongin-si, Gyeonggi-do 17104, Republic of Korea}

\nocollaboration{9}

%% Note that the \and command from previous versions of AASTeX is now
%% depreciated in this version as it is no longer necessary. AASTeX 
%% automatically takes care of all commas and "and"s between authors names.

%% AASTeX 6.3 has the new \collaboration and \nocollaboration commands to
%% provide the collaboration status of a group of authors. These commands 
%% can be used either before or after the list of corresponding authors. The
%% argument for \collaboration is the collaboration identifier. Authors are
%% encouraged to surround collaboration identifiers with ()s. The 
%% \nocollaboration command takes no argument and exists to indicate that
%% the nearby authors are not part of surrounding collaborations.

%% Mark off the abstract in the ``abstract'' environment. 
\begin{abstract}
Stellar kinematics is a powerful tool for understanding 
the formation process of stellar associations. Here, we present 
a kinematic study of the young stellar population in the Rosette 
nebula using the recent Gaia data and high-resolution spectra. 
We first isolate member candidates using the published mid-infrared 
photometric data and the list of X-ray sources. A total of 403 stars 
with similar parallaxes and proper motions are finally selected 
as members. The spatial distribution of the members shows that 
this star-forming region is highly substructured. The young open 
cluster NGC 2244 in the center of the nebula has a pattern of 
radial expansion and rotation. We discuss its implication on the 
cluster formation, e.g., monolithic cold collapse or hierarchical 
assembly. On the other hand, we also investigate three groups located 
around the border of the H {\scriptsize \textsc{II}} bubble. 
The western group seems to be spatially correlated with the adjacent 
gas structure, but their kinematics is not associated with that of the gas. 
The southern group does not show any systematic motion relative 
to NGC 2244. These two groups might be spontaneously formed 
in filaments of a turbulent cloud. The eastern group is spatially and 
kinematically associated with the gas pillar receding away from NGC 2244. 
This group might be formed by feedback from massive stars in NGC 2244. 
Our results suggest that the stellar population in the Rosette Nebula 
may form through three different processes: the expansion of stellar clusters, 
hierarchical star formation in turbulent clouds, and feedback-driven star 
formation.
\end{abstract}

%% Keywords should appear after the \end{abstract} command. 
%% See the online documentation for the full list of available subject
%% keywords and the rules for their use.
\keywords{Star formation (1569); Stellar kinematics (1608); Stellar 
associations (1582); Stellar dynamics (1596); Open star clusters (1160)}

%% From the front matter, we move on to the body of the paper.
%% Sections are demarcated by \section and \subsection, respectively.
%% Observe the use of the LaTeX \label
%% command after the \subsection to give a symbolic KEY to the
%% subsection for cross-referencing in a \ref command.
%% You can use LaTeX's \ref and \label commands to keep track of
%% cross-references to sections, equations, tables, and figures.
%% That way, if you change the order of any elements, LaTeX will
%% automatically renumber them.
%%
%% We recommend that authors also use the natbib \citep
%% and \citet commands to identify citations.  The citations are
%% tied to the reference list via symbolic KEYs. The KEY corresponds
%% to the KEY in the \bibitem in the reference list below. 

\section{Introduction} \label{sec:sec1}
Star formation hierarchically takes place on various spatial 
scales \citep{EEPZ00,G18}. Stellar complexes are the largest 
components of galactic superstructures spanning hundreds 
of parsecs. These complexes are composed of 
several stellar associations that extend up to several 
tens of parsecs. Single or multiple stellar clusters 
(a few parsecs) and a distributed stellar population constitute these 
stellar associations \citep{B64,KLB12}. The formation 
of these components is thought to be physically interconnected 
in space and time. In this context, stellar associations would be 
basic units of star formation in galaxies. Indeed, the majority 
of stars tend to form in such stellar systems \citep{LL03,PCA03}, 
and their initial mass functions are very similar to that 
derived from field stars in the Galaxy \citep{MS78,BPS07}.  

Stellar associations are gravitationally unbound and 
substructured \citep{MD17,KHS19,LNGR19,LHY20}. The internal 
structure and kinematics are the keys to understanding 
their formation process. A classical model 
attempted to explain the observed structures and unboundness 
of OB associations in the context of dynamical evolution of 
embedded clusters. Once embedded clusters have formed 
in giant molecular clouds, they begin to expand after rapid gas 
expulsion by winds and outflows of embedded OB stars 
\citep{T78,H80,LMD84,KAH01,BK13,BK15}. \citet{KAH01} 
performed $N$-body simulations for model clusters with the 
properties similar to those of the Orion Nebula Cluster that were 
observationally constrained. As a result, they found that only about 
30 \% of the initial cluster members remain in the model clusters. 
Their results imply that open clusters can be the cores of 
unbound OB associations. 

The structural features observed in star-forming complexes 
are similar to those found in molecular clouds \citep{DHM90,
EEPZ00,A15}. It has been suggested that turbulence is responsible 
for the origin of these substructures in molecular clouds 
\citep{L81,PJGN01}. Density fluctuation caused by turbulence 
results in locally different star formation efficiency, and 
therefore groups of stars with different stellar densities 
form along the substructures on short timescales \citep{BSCB11,K12}. 

On the other hand, massive stars can play crucial roles in 
regulating star formation in either constructive or destructive
way. Far-ultraviolet radiation and stellar wind from massive 
stars disperse the remaining gas in their close neighborhood, 
lowering star formation efficiency and eventually terminating star formation 
\citep{DEB12,DEB13}. In contrast, expanding H {\scriptsize \textsc{II}} 
bubbles driven by massive stars can compress the 
surrounding material, triggering the formation of new 
generations of stars further away from the massive stars 
\citep{EL77}. OB associations could also form through this 
self-propagating star formation.

Explaining the formation of stellar associations from 
observed features seems complicated and impossible 
by one specific model. The signature of dynamical evolution 
of central stellar clusters has been reported. For example, a pattern 
of expansion has been detected in young stellar clusters 
within many associations \citep{KCS18,C-G19,LNGR19,KHS19}. 
Recently, a kinematic study showed that a distributed 
stellar population spread over 20 pc in W4 can be formed 
by cluster members radially escaping from the central cluster 
IC 1805 \citep{LHY20}. In addition, even larger scale expansion 
of associations spanning several hundreds of parsecs toward the 
Perseus arm was also reported \citep{R-Z19,DVT21}.

Several stellar groups were found at the border of the 
W4 H {\scriptsize \textsc{II}} region \citep{PSP19}, and 
their projected distance from IC 1805 cannot be explained 
by crossing time of the escaping stars \citep{LHY20}. Instead, feedback-driven star 
formation may be a possible explanation for the formation 
of these groups. Many signposts of feedback-driven 
star formation have been steadily reported in various star-forming 
regions (SFRs) \citep[etc]{FHS02,SHB04,KAG08}. For instance, a 
stellar group in NGC 1893 is located in the vicinity of the gas 
pillar Sim 130 \citep{MN02,SPO07}. This group is younger 
than the central cluster containing massive O-type stars 
\citep{LSK14} and receding away from the cluster at a similar 
velocity to that of the gas pillar \citep{LSB18}. 

\begin{figure*}[t]
\includegraphics[width=7.5cm]{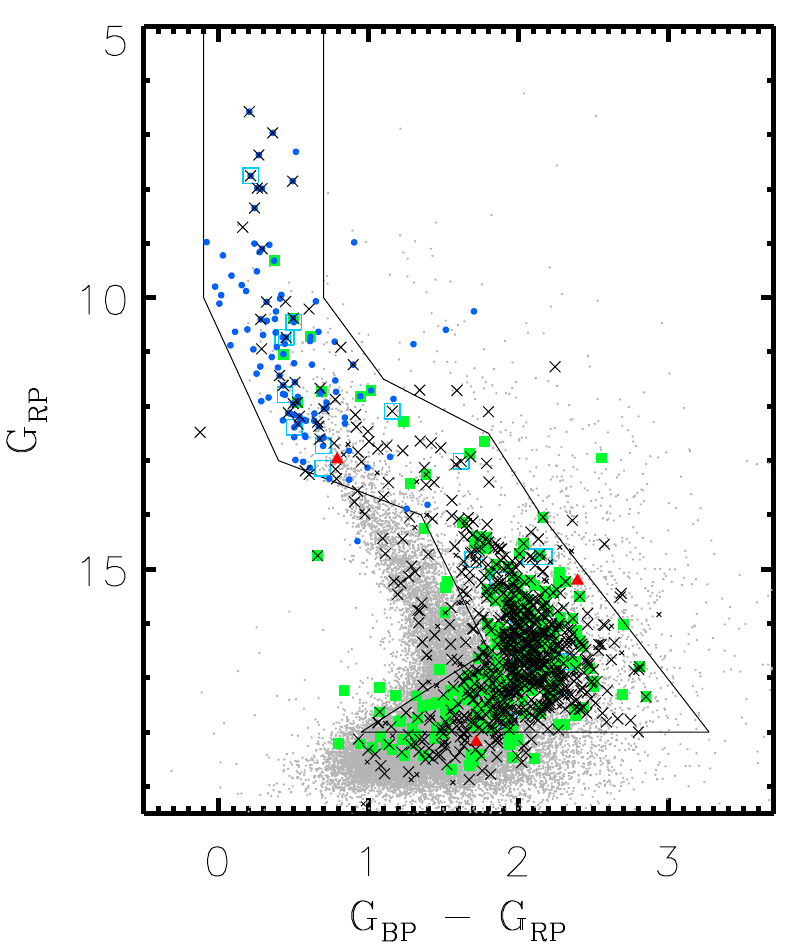}\includegraphics[width=9.2cm]{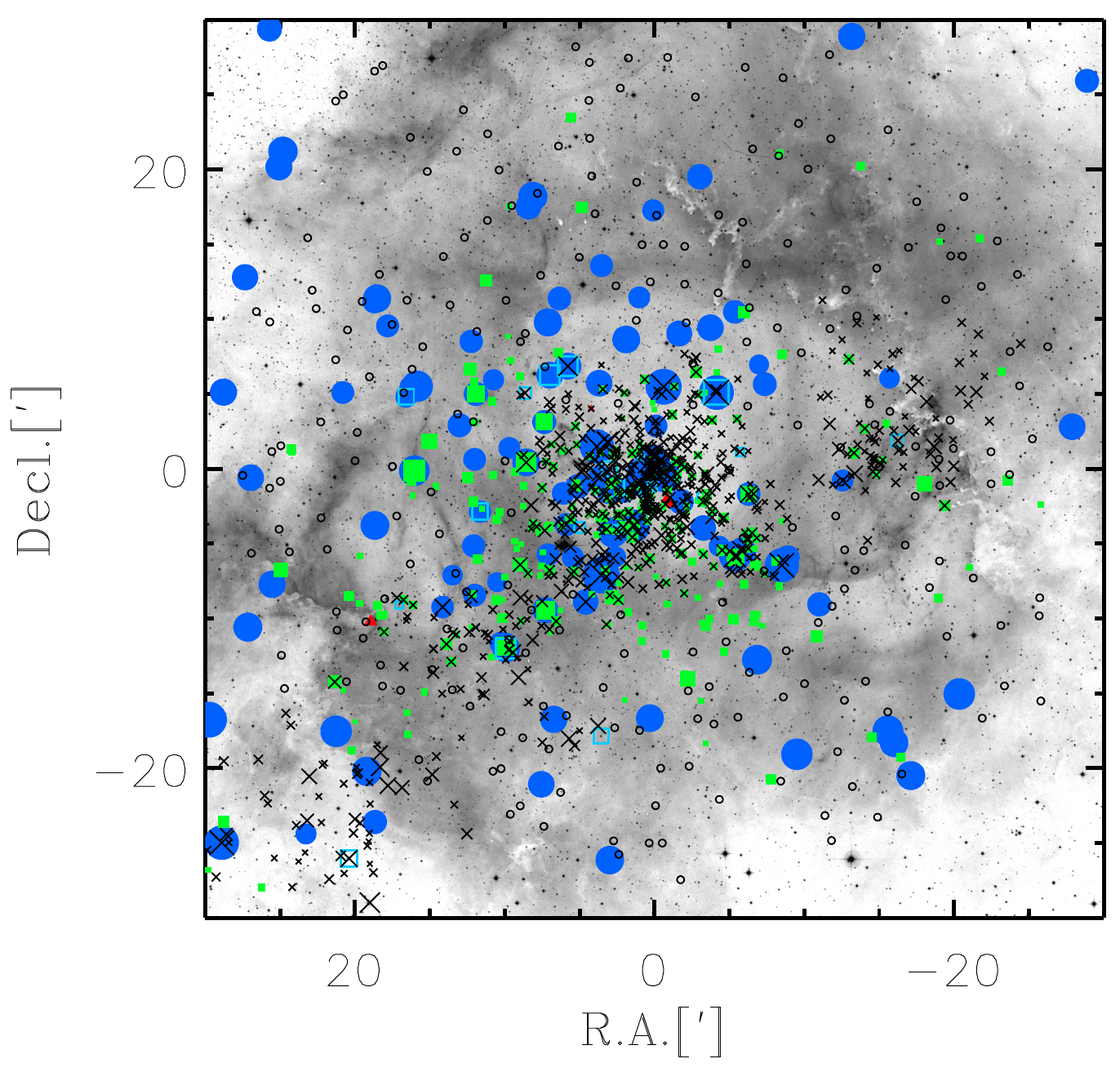}
\caption{CMD (left) and spatial distribution (right) 
of member candidates in the Rosette Nebula. The photometric data were taken 
from the Gaia Data Release 2 \citep{gdr2}. Blue dots, red triangles, green filled 
square, cyan open square, black large cross, small cross, and grey small dots 
represent early-type star (O- or B-type), Class I, Class II, object with a 
transitional disk, X-ray source, X-ray source candidates, and all stars in the survey 
region, respectively. We set a locus (black solid line) that contains the majority of member 
candidates. In the right panel, the fiber positions (black open circles) are superimposed on the optical image 
taken from the Digitized Sky Survey. The positions of stars are relative to the coordinate 
R.A. = 06$^{\mathrm{h}}$ 31$^{\mathrm{m}}$ $55\fs00$, Decl. = 
$+04^{\circ}$ 56$^{\prime}$ $30\farcs0$ (J2000). The size of the symbols is 
proportional to the brightness of individual stars. }\label{fig1}
\end{figure*}

The signatures of hierarchical star formation in turbulent clouds 
have been investigated using stellar proper motions (PMs) under 
the assumption that the age of OB associations is young enough 
to preserve the kinematic properties of their natal clouds just 
before onset of star formation. For example, the Cygnus OB2 and Carina OB1 
associations are composed of spatially and kinematically distinct 
subgroups \citep{LNGR19}. In addition, a correlation between 
the size and two-dimensional velocity dispersion of these 
subgroups was found, which has a similar power law index 
to that found in molecular clouds \citep{L81}. These first observational 
results were supported by an extensive kinematic study on 
larger samples of the Galactic OB associations \citep{WKR20}.  

Hence, we hypothesize that stellar associations may form 
through three different processes, i.e., dynamical evolution of 
stellar clusters, feedback from massive stars, and star formation 
in turbulent clouds. In this context, the Rosette Nebula, the 
most active SFR in the Monoceros OB2 association, is an ideal 
site to examine the hypothesis because this region is composed 
of several stellar groups in a range of 30 pc. The open cluster NGC 2244 
is centered at the cavity of the Rosette Nebula and contains several 
tens of OB stars \citep{RL08,MNR09} as well as a large number 
of low-mass young stellar objects (YSOs) \citep{BMR07,WTF08,MSR19}. 
These massive stars may be the main ionizing sources of the 
H {\scriptsize \textsc{II}} region. Additional, smaller, stellar groups 
are found at the border of the nebula \citep{L05,WFT09,WFT10}. 

The Rosette Molecular Cloud is located to the east of the H 
{\scriptsize \textsc{II}} region, and star-forming activity is 
still ongoing there. \citet{PL97} reported the presence of 
seven embedded clusters. Later, \citet{REFL08} identified 
four more clusters (see also the historical review on this 
SFR by \citealt{RL08}). Extensive {\it Spitzer} and {\it Chandra} 
surveys continuously discovered deeply embedded clusters 
composed of active YSOs \citep{PRG08,WFT09,CMF13}. \citet{YLR13} 
investigated the star formation history on different scales 
from clusters to the molecular cloud and claimed that 
star formation occurred almost simultaneously across 
the cloud. The relation between the YSO ratio and 
extinction that they found suggests that the densest regions in the 
cloud are the favorable sites of star formation and that gas 
is rapidly evacuating out, lowering star formation efficiencies 
in the cloud.

In this study, we aim to understand the formation process 
of the stellar groups in the Rosette Nebula using gas and 
stellar kinematics. The splendid performance of the Gaia 
mission \citep{gaia16} opened a new window to study 
such young stellar systems. The Gaia astrometric data \citep{gedr3} 
provide a better understanding of the kinematic properties 
of stellar groups in this SFR when combined with spectroscopic 
data of stars and gas. Observations and data that we used are described 
in Section~\ref{sec:sec2}. In Section~\ref{sec:sec3}, the 
method of member selection is addressed. We investigate 
the motions of stars using PMs \citep{gedr3} and radial 
velocities (RVs) in Section~\ref{sec:sec4}. The 
physical associations between stars and gas are 
also probed. The formation process of NGC 2244 
and stellar groups around the cluster is discussed in 
Section~\ref{sec:sec5}. Finally, we summarize our results 
in Section~\ref{sec:sec6}.  

\section{Data} \label{sec:sec2}
\subsection{Selection of member candidates}
Our targets are young stars within an $1^{\circ} \times 1^{\circ}$ 
region centered at R.A. = 06$^{\mathrm{h}}$ 31$^{\mathrm{m}}$ 
$55\fs00$, Decl. = $+04^{\circ}$ 56$^{\prime}$ $30\farcs0$ (J2000). 
This SFR ($b \sim -2\fdg072$) is located in the Galactic plane, and 
therefore a large number of field interlopers are also 
observed in the same field of view. We intend to 
isolate member candidates first using multiple sets of archival data 
in this section, and then the final decision on membership is made 
by using the recent astrometric data from the Gaia Early Data Release 3 
(EDR3, \citealt{gedr3}) in Section~\ref{sec:sec3}.

To identify probable member candidates in the survey 
region, we considered the intrinsic properties 
of young stars. O- and B-types stars spread over this SFR 
are probable member candidates because the positions 
of these stars are close to birthplaces within their short 
lifetime, particularly for O-type stars. We took a list of OB 
stars from the data bases of MK classifications 
\citep[SIMBAD]{R03,MP03,S09,MSM13} and combined them 
into a list after checking for duplicates. A total of 112 OB 
stars were selected as member candidates.

A large number of low-mass YSOs have a warm 
circumstellar disk. These disk-bearing YSOs appear 
bright in infrared passbands \citep{L87,GMM08,KLB12}. YSO 
candidates were identified using the published {\it Spitzer} 
and AllWISE data \citep{BMR07,C13} (see Appendix for 
detail). We found 291 YSO candidates (10 Class I, 262 Class II, 
and 19 YSOs with a transitional disk) from the {\it Spitzer} 
data in total, while a total of 84 candidates (six Class I, 
76 Class II, and two objects with a transitional disk) were 
identified from the AllWISE data. There are 37 YSO 
candidates in common between these data sets. Note that 
the survey region ($30^{\prime} \times 30^{\prime}$) 
by {\it Spitzer} \citep{BMR07} is smaller than ours. Our 
list contains 338 YSO candidates in total. 

We used photometric data from the Gaia Data Release 
2 (DR2,  \citealt{gdr2}) instead of the recent EDR3 
\citep{gedr3} because the reddening law is empirically well 
established \citep{WC19} for the former. We found 
counterparts for all the OB stars and for 320 YSO 
candidates in the Gaia DR2 \citep{gdr2}. In addition, 
we searched for counterparts of X-ray sources detected in this SFR 
\citep{WTF08,WFT09,WFT10}. Stars found 
within searching radii of $1\farcs0$ and $1\farcs5$ were 
classified as X-ray sources and X-ray source candidates, 
respectively. A total of 720 X-ray sources 
and 134 candidates were found in the Gaia data. 
Note that the lists of X-ray sources are complete 
down to 0.5 -- 1 M$_{\sun}$ \citep{WTF08,WFT09,WFT10}.

\begin{table*}
\begin{center}
\setlength\tabcolsep{3pt}
\caption{Summary of observations \label{tab1}}
\begin{tabular}{llcccccc}
\tableline\tableline\scriptsize
 & Date & Setup & Filter &  $N\tablenotemark{1}$ & Exposure time & Binning & Seeing\\
\tableline 
          &2019 October  13      &   1  & RV31  &     111  & 35 min. $\times$ 3  &  $2\times2$  &  $0\farcs9$\\    
         &2019 November 13    &   2  & RV31  &    110  & 40 min. $\times$ 3   &   $2\times2$ &  $1\farcs4$\\
YSOs &                                 &       &           &                   &  38 min. $\times$ 1  &                      &                  \\
          &                       &   3  & RV31  &       95  & 37 min. $\times$ 1   &   $2\times2$ &    $1\farcs3$\\
          &                       &       &           &                   &  35 min. $\times$ 1  &                      &                    \\
         &2019 November 18   &   3  &  RV31 &       95  & 35 min. $\times$ 2    &  $2\times2$ &    $1\farcs3$\\
\tableline
Ionized gas & 2019 November  9   &  4    & OB25 &     238 & 15 min. $\times$ 5  &  $1\times1$   &    $0\farcs9$ \\
\tableline
\end{tabular}
\tablenotetext{1}{$N$ represents the number of observed targets.}
\end{center}
\end{table*}

Figure~\ref{fig1} displays the color-magnitude diagram (CMD) of 
stars in the Rosette Nebula. Most member candidates seem to occupy 
a specific locus in the diagram. The main-sequence turn-on 
is found at about 13 mag in $G_{RP}$. There are several B-type 
stars fainter than YSOs near the main-sequence turn-on. These 
stars may be background early-type stars. A large fraction of 
the X-ray sources may be YSO candidates, while some 
X-ray sources appear to fall on the main-sequence band. 
These X-ray emitting stars may be late-type field stars. All the 
YSO candidates with infrared excess emission were considered as member 
candidates because their colors and magnitudes could vary with 
different levels of internal extinction by material in disks or 
envelopes and accretion activities \citep{LJL20}. Photometric 
errors increase with brightness, and therefore we limited our 
sample to 840 candidates brighter than 18 mag. The selection 
criteria of member candidates can be summarized as below:

\begin{enumerate}
\item $G_{RP} < 18$ mag
\item O- or B-type stars brighter than 13.5 mag in $G_{RP}$
\item Class I YSOs
\item Class II YSOs
\item YSOs with a transitional disk
\item X-ray sources within the locus (see Figure~\ref{fig1}).
\end{enumerate}

\subsection{Radial velocities}
We performed spectroscopic observations of 316 YSO candidates with 
or without an X-ray emission and ionized gas on 2019 October 13, November 9, 13, and 18 
using the high-resolution ($R \sim 34000$) multi-object 
spectrograph {\it Hectochelle} \citep{SFC11} on the 6.5-m telescope 
of the MMT observatory. The spectra of the YSO candidates 
were taken with the RV31 filter (5150 -- 5300 \AA) in $2\times2$ 
binning mode to achieve good signal-to-noise ratios. In addition, 
several tens of fibers were simultaneously assigned to blank sky 
to obtain sky spectra. The spectra of ionized gas were obtained 
from the fibers assigned to 238 positions on the Rosette Nebula 
(see Figure~\ref{fig1}). The oder separating filter OB25 (6475 -- 6630 \AA) 
was used to observe the forbidden line [N {\scriptsize \textsc{II}}] 
$\lambda6584$. For calibration, dome flat and ThAr 
lamp spectra were also obtained just before and after the target 
observation. We present a summary of our observations in 
Table~\ref{tab1}.

We preprocessed the raw mosaic frames using the IRAF\footnote{Image 
Reduction and Analysis Facility is developed and distributed by the 
National Optical Astronomy Observatories, which is operated by the 
Association of Universities for Research in Astronomy under operative 
agreement with the National Science Foundation.}/{\tt MSCRED} 
packages in a standard way. One-dimensional spectra 
were then extracted from the reduced frames using the 
{\tt dofiber} task in the IRAF/{\tt SPECRED} package. Target 
spectra were flattened using dome flat spectra. The solutions 
for the wavelength calibration obtained from ThAr spectra 
were applied to the target and sky spectra.

Our observations were performed under bright sky condition, and 
therefore the scattered light was unevenly illuminated over 
the field of view ($1^{\circ}$ in diameter), resulting in a spatial 
variation of sky levels. For a given observing setup, we constructed 
a map of the sky levels and combined sky spectra into 
a master sky spectrum with an improved signal-to-noise ratio. 
The sky level at a target position was inferred by interpolating 
the position of the target to the map, and its sky spectrum was 
obtained from the master sky spectrum scaled to the inferred 
sky level. For each target spectrum, we subtracted the corresponding 
scaled master sky spectra and then combined sky-subtracted spectra 
into a single spectrum for the same target observed on the same night. Finally, the target 
spectra were normalized by using continuum levels traced from 
a cubic spline interpolation. 

Since the spectra of low-mass stars contain a large number of 
metallic lines, their RVs can be measured more precisely than 
those of high-mass stars. The RVs of the YSO candidates were 
measured by applying a cross-correlation technique to the 
observed spectra. Assuming the Solar chemical abundance, a total of 31 
synthetic spectra in a wide temperature range of 3800--9880 K 
were generated from the \textsc{MOOG} code and Kurucz 
ODFNEW model stellar atmosphere \citep{S73,CK04}. 
We derived cross-correlation function (CCF) between 
the observed spectra and synthetic ones using the {\tt xcsao} 
task in the \textsc{RVSAO} package \citep{KM98} and 
adopted the central velocities at the strongest CCF peaks 
as RVs. The errors on RVs were estimated 
from the equation $3w/8(1+h/\sqrt{2}\sigma_a)$, where $w$, $h$, 
and $\sigma_a$ represent the full widths at half-maximum 
of CCFs, their amplitudes, and the rms from 
antisymmetric components, respectively \citep{TD79,KM98}. 
The measured RVs were converted to velocities in the local 
standard of rest frame using the \textsc{IRAF}/{\tt RVCORRECT} task. 

The spectra of 95 YSO candidates (Setup 3 in Table~\ref{tab1}) 
were obtained on two different nights. The RVs of these 
stars were measured for each night. The median difference between 
two RV measurements for 52 common stars is about 0.3 km 
s$^{-1}$, which is smaller than the mean of measurement 
errors ($\sim$ 1.5 km s$^{-1}$). The weighted means of 
two measurements were adopted as the RVs of given stars, where 
the inverse of the squared error was used as the weight value. 
We measured the RVs of 224 out of 316 YSO candidates in total. 
The spectra of the other 92 stars have insufficient signals to 
derive CCFs, or some of them show only emission lines.

\begin{figure*}[t]
\epsscale{1.0}
\plotone{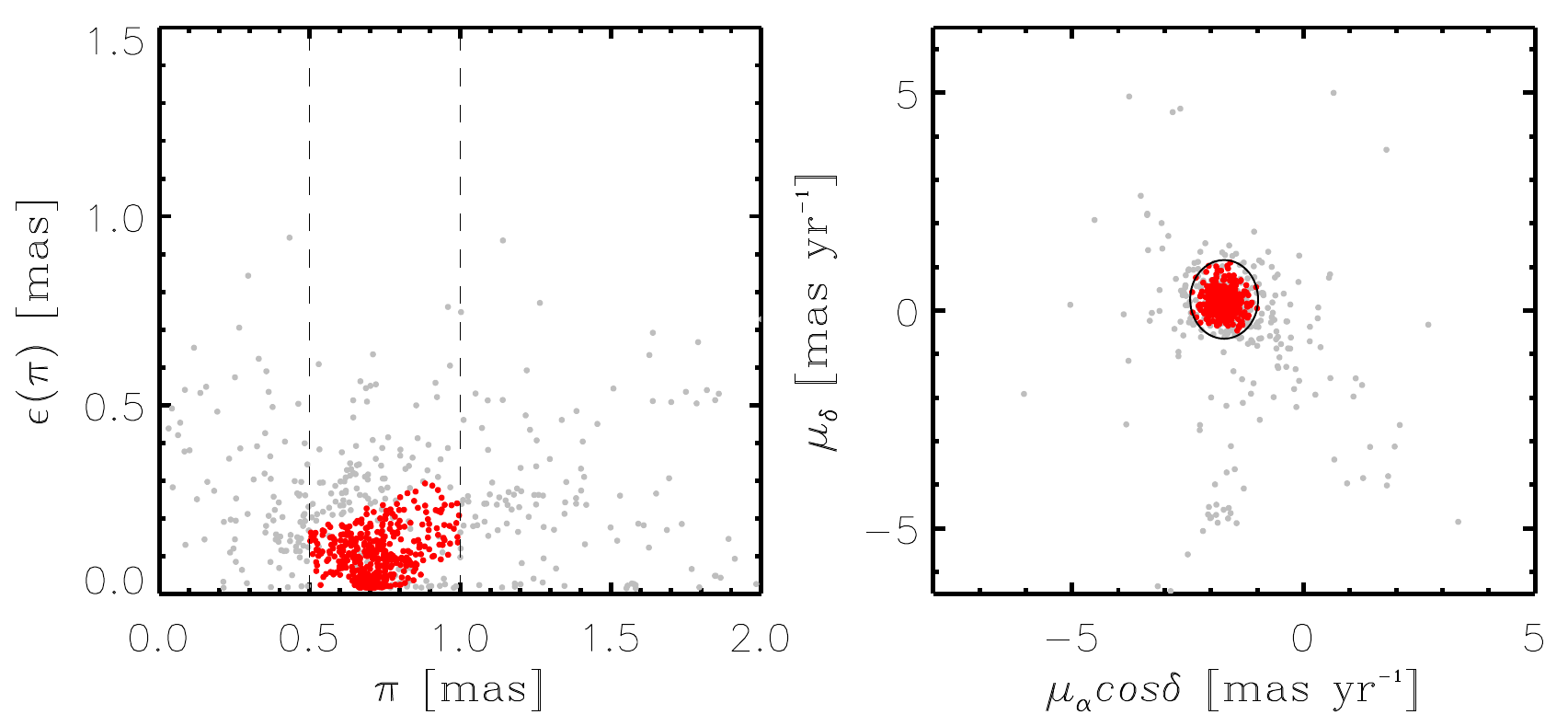}
\caption{Parallax (left) and PM (right) distributions of 
member candidates (grey dots). Dashed lines in the left panel represent the lower 
and upper limits in parallax for selecting genuine members. The 
parallaxes from the Gaia EDR3 \citep{gedr3} are corrected for 
zero-point offsets \citep{LBB20}. The ellipse in the 
right panel shows the region confined to 3.5 times the 
standard deviation from the weighted mean PMs, where the 
inverse of the squared PM error was used as the weight value. 
The selected members were plotted by red dots.}\label{fig2}
\end{figure*}

We used the forbidden line [N {\scriptsize \textsc{II}}] 
$\lambda$6584 to probe the kinematics of ionized gas. In 
general, the critical density of this line is higher than the 
typical electron densities of the Galactic H {\scriptsize \textsc{II}} 
regions \citep{CMSC00}, and photons emitted from the singly ionized 
nitrogen atoms are not absorbed along the line of sight. Therefore, 
this emission line can be used to trace the structure and kinematics 
of ionized gas distributed across an H {\scriptsize \textsc{II}} 
region. The RVs of the ionized gas were measured from the line 
center of the best-fit Gaussian profile. We fit multiple Gaussian 
profiles to some complex line profiles and measured the RVs of the 
multiple components along the line of sight.

\subsection{$^{12}$CO ($J = 1-0$) and $^{13}$CO ($J=1-0$) data} 
We used radio data to investigate the physical association between 
stellar groups and remaining molecular gas. The $^{12}$CO and 
$^{13}$CO ($J = 1-0$) line maps were obtained from \citet{HWB06}. 
These radio maps covering both the Rosette Nebula and 
the eastern molecular cloud are useful for investigating the velocity fields 
of remaining molecular gas at the boundary of the H {\scriptsize \textsc{II}} region. 
In addition, the excitation temperature $T_{\mathrm{ex}}$ and column densities 
can be estimated by assuming of local thermodynamic equilibrium. 

Since the $^{12}$CO line is, in general, optically thick, $T_{\mathrm{ex}}$ can 
be derived from the following equation by \citet{PCG08}: 
\begin{equation}
T_{\mathrm{ex}} = {5.5 \mathrm{K} \over \ln(1+5.5 \mathrm{K}/T_{\max}(^{12}\mathrm{CO}) + 0.82 \mathrm{K})}
\end{equation}
\noindent where $T_{\max}(^{12}\mathrm{CO})$ is the brightness 
temperature at the peak intensity of $^{12}$CO. The column density 
of $^{13}$CO was estimated from the integrated intensity along 
the line of sight (in unit of K km s$^{-1}$) assuming an optically 
thin case \citep{PCG08}. On the other hand, the column density 
of $^{12}$CO was estimated using the relation of \citet{SSS86}. 
However, this relation considers an optically thin case. The effects 
of optical depth were corrected by adopting the correction factor 
addressed in \citet{SCG07}. 

The integrated intensity maps of molecular gas were also 
constructed from the $^{12}$CO $(J=1-0)$ cube data. Note that 
the $^{12}$CO lines between velocities of $-5$ to 25 km s$^{-1}$ 
were integrated. In addition, position-velocity diagrams were 
obtained by integrating the data cube along R.A. and declination. 

\begin{figure}[t]
\includegraphics[width=7.1cm]{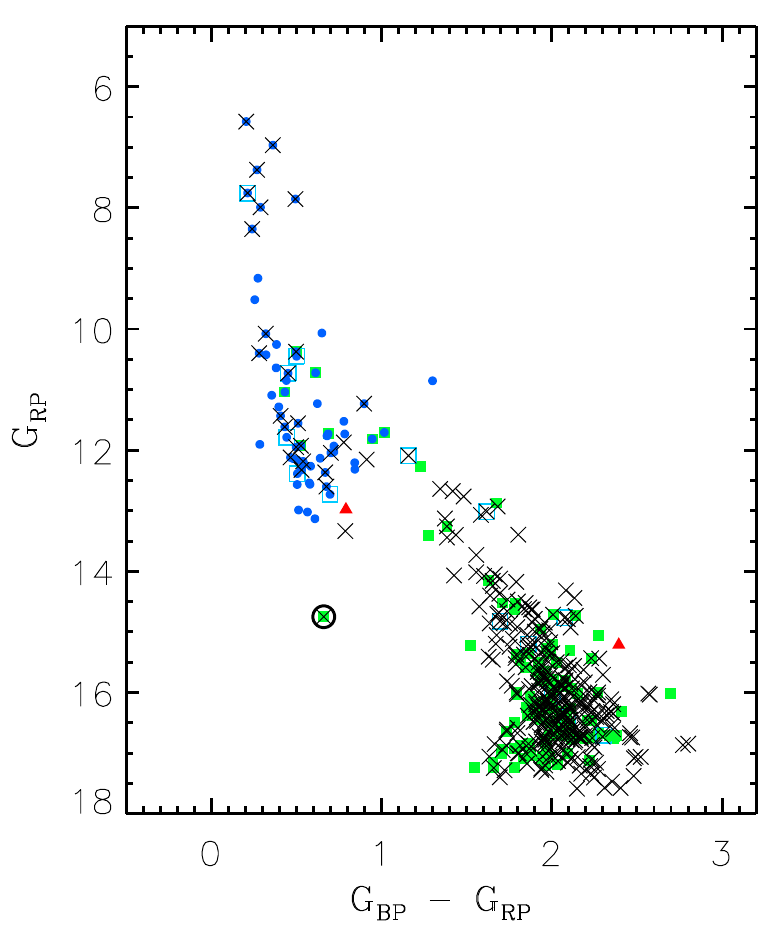}
\includegraphics[width=7cm]{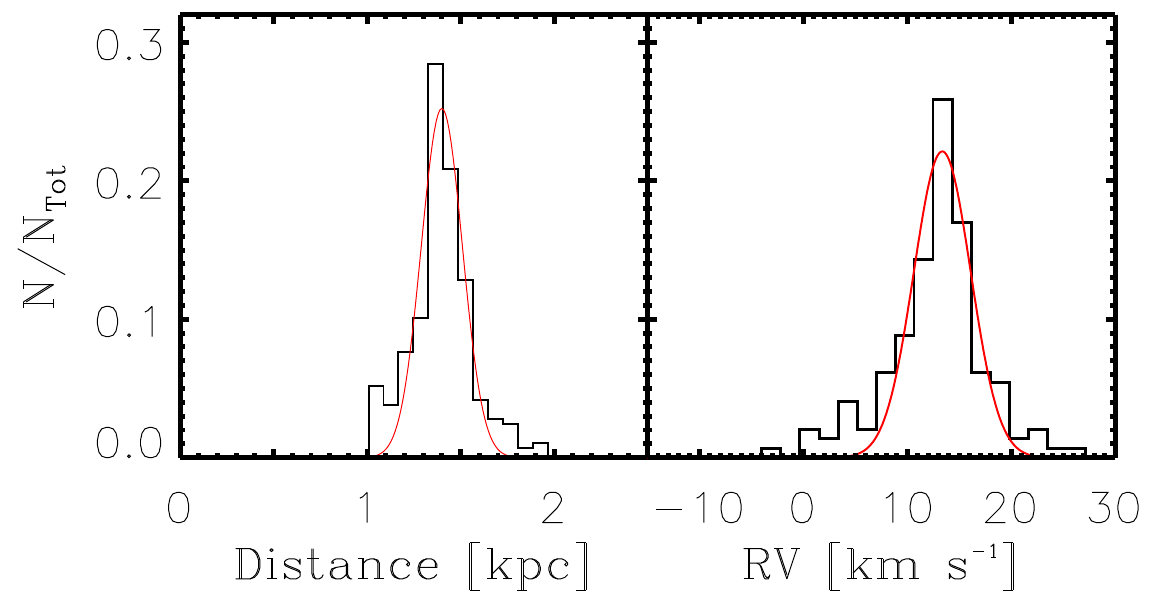}
\caption{Selected members. The upper panel 
displays the CMD of the members taken from the 
photometric data of the Gaia DR2 \citep{gdr2}. An open circle 
denotes a nonmember with an X-ray emission 
(see the main text for detail). The other symbols 
are the same as Figure~\ref{fig1}. 
The lower-left and lower-right panels show 
the distance and RV distributions of the members, respectively. 
The distances were computed from the inversion 
of the recent Gaia parallaxes after correction for 
zero-point offsets \citep{gedr3,LBB20}. 
Red curves represent the best-fit Gaussian distributions. }\label{fig3}
\end{figure}

\begin{deluxetable}{lccccccccccccccccccccc}
\rotate
\setlength\tabcolsep{1.5pt}
\tabletypesize{\tiny}
\tablewidth{0pt}
\tablecaption{List of members. \label{tab2}}
\tablehead{\colhead{Sq.} & \colhead{R.A. (2000)} & \colhead{Dec. (2000)} & \colhead{$\pi$} & \colhead{$\epsilon(\pi)$} &
\colhead{$\mu_{\alpha}\cos\delta$}  & \colhead{$\epsilon(\mu_{\alpha}\cos\delta)$} &
\colhead{$\mu_{\delta}$} & \colhead{$\epsilon(\mu_{\delta})$} & \colhead{$G$} & \colhead{$\epsilon(G)$} &
\colhead{$G_{BP}$} & \colhead{$\epsilon(G_{BP})$} & \colhead{$G_{RP}$} & \colhead{$\epsilon(G_{RP})$} & \colhead{$G_{BP}-G_{RP}$} & \colhead{Early-type} &
\colhead{YSO} & \colhead{X-ray} &  \colhead{RV} & \colhead{$\epsilon$(RV)} & \colhead{Binarity flag} \\
 & \colhead{(h:m:s)} & \colhead{($\degr:\arcmin:\arcsec$)} &  \colhead{(mas)} & \colhead{(mas)} & \colhead{(mas yr$^{-1}$)} &\colhead{(mas yr$^{-1}$)} &
\colhead{(mas yr$^{-1}$)} &\colhead{(mas yr$^{-1}$)} & \colhead(mag) & \colhead(mag) & \colhead(mag) & \colhead(mag) & \colhead(mag) & \colhead(mag) &
\colhead(mag) & & & & \colhead{(km s$^{-1}$)} & \colhead{(km s$^{-1}$)} &  }
\startdata
  1 & 06:30:16.81 & +05:04:51.9 &     0.6459 &  0.0975 &    -1.975 &  0.101 &     0.159 &  0.086 & 17.5846 & 0.0102 & 18.4017 & 0.0792 & 16.3339 & 0.0241 &  2.0677 &   &   & X &  \nodata &\nodata &     \\
  2 & 06:30:26.49 & +05:02:41.1 &     0.9872 &  0.1871 &    -2.220 &  0.216 &     0.014 &  0.173 & 18.6931 & 0.0044 & 19.2147 & 0.0372 & 17.2773 & 0.0173 &  1.9374 &   &   & X &  \nodata &\nodata &     \\
  3 & 06:30:28.15 & +05:05:30.3 &     0.6576 &  0.1224 &    -2.267 &  0.130 &     0.110 &  0.127 & 17.4778 & 0.0033 & 18.3454 & 0.0266 & 16.2546 & 0.0112 &  2.0908 &   &   & X &  \nodata &\nodata &     \\
  4 & 06:30:30.87 & +05:00:49.3 &     0.5226 &  0.1168 &    -1.855 &  0.129 &    -0.044 &  0.104 & 17.6367 & 0.0019 & 18.7096 & 0.0176 & 16.3491 & 0.0070 &  2.3604 &   &   & X &   16.446 &  4.037 &     \\
  5 & 06:30:32.61 & +05:06:16.6 &     0.9460 &  0.1443 &    -1.765 &  0.167 &    -0.228 &  0.152 & 17.9478 & 0.0081 & 18.7829 & 0.0303 & 16.5876 & 0.0066 &  2.1953 &   &   & X &  \nodata &\nodata &     \\
  6 & 06:30:41.00 & +04:57:30.9 &     0.7264 &  0.0720 &    -1.737 &  0.083 &    -0.214 &  0.067 & 16.9469 & 0.0081 & 17.8253 & 0.0296 & 15.7647 & 0.0251 &  2.0606 &   &   & X &    6.683 &  0.605 &     \\
  7 & 06:30:42.01 & +05:00:59.7 &     0.7461 &  0.0296 &    -1.713 &  0.034 &    -0.168 &  0.027 & 15.0324 & 0.0009 & 15.8159 & 0.0053 & 14.1195 & 0.0032 &  1.6965 &   &   & X &   18.709 &  0.468 &     \\
  8 & 06:30:44.15 & +05:02:21.0 &     0.6525 &  0.0871 &    -1.957 &  0.092 &     0.223 &  0.075 & 17.0066 & 0.0043 & 17.9700 & 0.0145 & 15.8613 & 0.0107 &  2.1088 &   &   & X &   17.468 &  0.999 &     \\
  9 & 06:30:46.18 & +05:01:51.3 &     0.7509 &  0.0720 &    -1.849 &  0.082 &     0.222 &  0.064 & 16.5845 & 0.0022 & 17.7197 & 0.0176 & 15.4415 & 0.0079 &  2.2782 &   &   & X &   21.966 &  4.375 &     \\
 10 & 06:30:46.77 & +04:59:04.2 &     0.5454 &  0.0441 &    -1.810 &  0.050 &    -0.108 &  0.040 & 15.8938 & 0.0028 & 16.7182 & 0.0105 & 14.9212 & 0.0069 &  1.7970 &   &   & X &   16.465 &  0.579 &     \\
\enddata
\tablecomments{Column (1) : Sequential number. Columns (2) and (3) : The equatorial coordinates of members. Columns (4) and (5) : Absolute parallax and its
standard error. Columns (6) and (7) : PM in the direction of right ascension and its standard error. Columns (8) and (9): PM in the direction
of declination and its standard error. Columns (10) and (11) : $G$ magnitude and its standard error. Columns (12) and (13) : $G_{BP}$ magnitude and
its standard error. Columns (14) and (15) : $G_{RP}$ magnitude and its standard error. Column (16) : $G_{BP} - G_{RP}$ color index. Column (17) : Early-type
members. `E' represents O- or B-type stars obtained from the data bases of MK classification \citep[SIMBAD]{R03,MP03,S09,MSM13}. Column (18) : YSO classification.
`1', `2', and `T' denote Class I, Class II, and YSOs with a transitional disk, respectively. Column (19) : X-ray source. 'X' and 'x' indicate X-ray source and
X-ray source candidates, respectively. Columns (20) and (21) : RV and its error. Columns (22) : Binarity flag. `SB2' represents
a double-lines spectroscopic binary candidate. The parallax and PM were taken from the Gaia Early Data Release 3 \citep{gedr3}, while the photometric data were
obtained from the Gaia Data Release 2 \citep{gdr2}. We corrected the zero-point offsets for the Gaia parallaxes according to the recipe of \citet{LBB20}. The full table is available electronically.}
\end{deluxetable}

\section{Member selection} \label{sec:sec3}
We selected member candidates based on the intrinsic properties 
of young stars, but some nonmembers with similar properties 
may be included in our candidate list. The astrometric data from 
the Gaia EDR3 \citep{gedr3} allow us to better isolate 
genuine members. \citet{LBB20} found the zero-point 
offset in parallax that depends on magnitude, color, and position. 
For reliable member selection, we corrected the zero-point 
offsets for the parallaxes of individual member candidates 
using the public python code \citep[\url{https://gitlab.com/icc-ub/public/gaiadr3_zeropoint}]{LBB20}\footnote{Note that using the global 
zero-point (0.017 mas) mentioned by \citet{LBB20} does 
not change the results in any significant way}.

Figure~\ref{fig2} displays the parallax and PM distributions of the 
member candidates. Stars with negative parallaxes or close companion 
(duplication flag = 1 or RUWE $>$ 1.4), or without astrometric 
parameters were not used in analysis. We first limited members to stars between 0.5 
and 1.0 mas in parallax, given the distance to NGC 2244 determined 
by previous studies (1.4 to 1.7 kpc; \citealt{OI81,PTW87,HPV00,PS02,MSR19}). 
Stars with parallax smaller than three times the associated 
error and PMs larger than 3.5 times the standard deviation from the 
weighted mean PMs were excluded. Note that the inverse of the 
squared PM errors was used as the weight value. We repeated the 
same procedure until the mean PMs and standard deviations 
converged to constant values. 

\begin{figure*}[t]
\includegraphics[width=9cm]{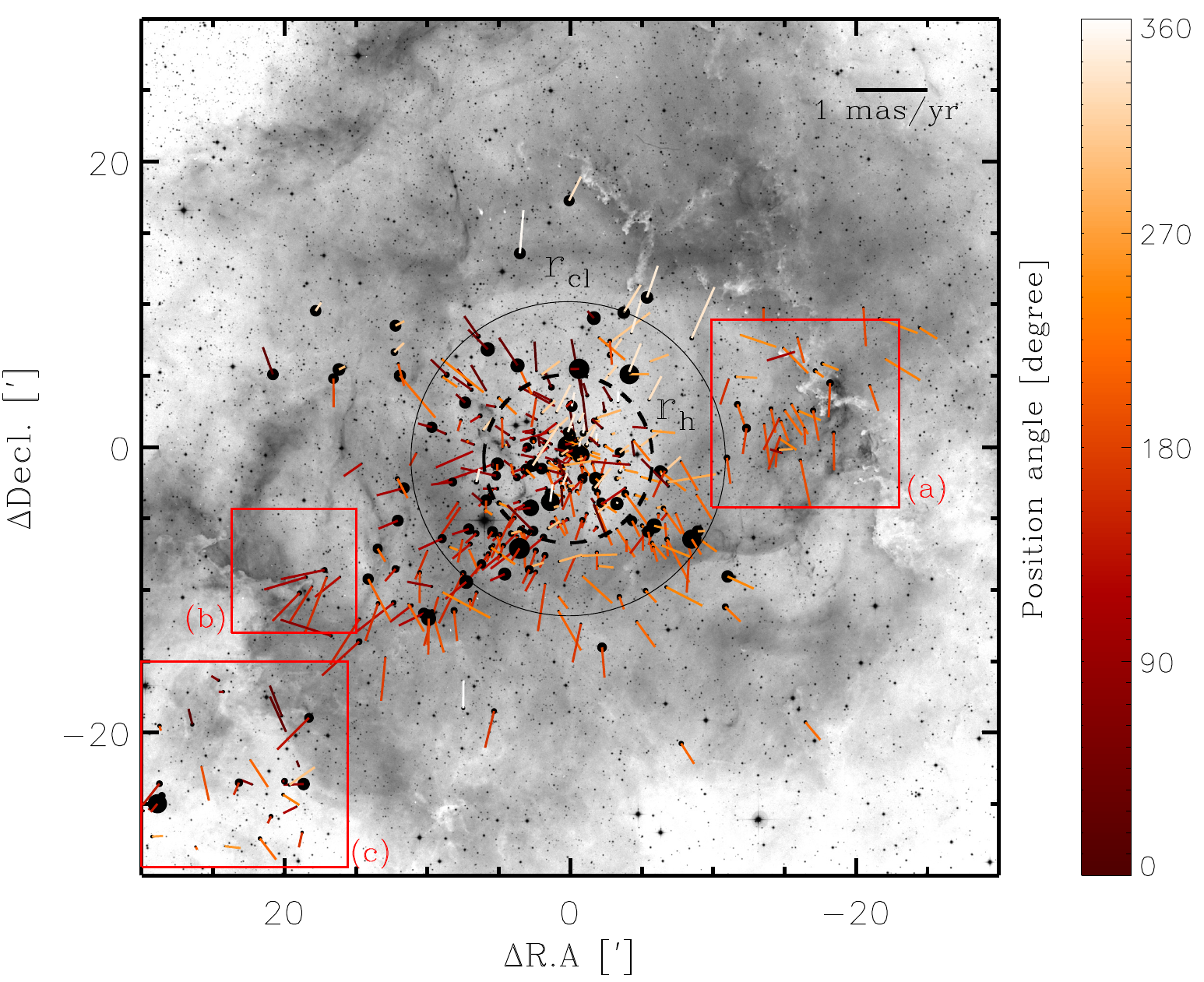}\includegraphics[width=7.5cm]{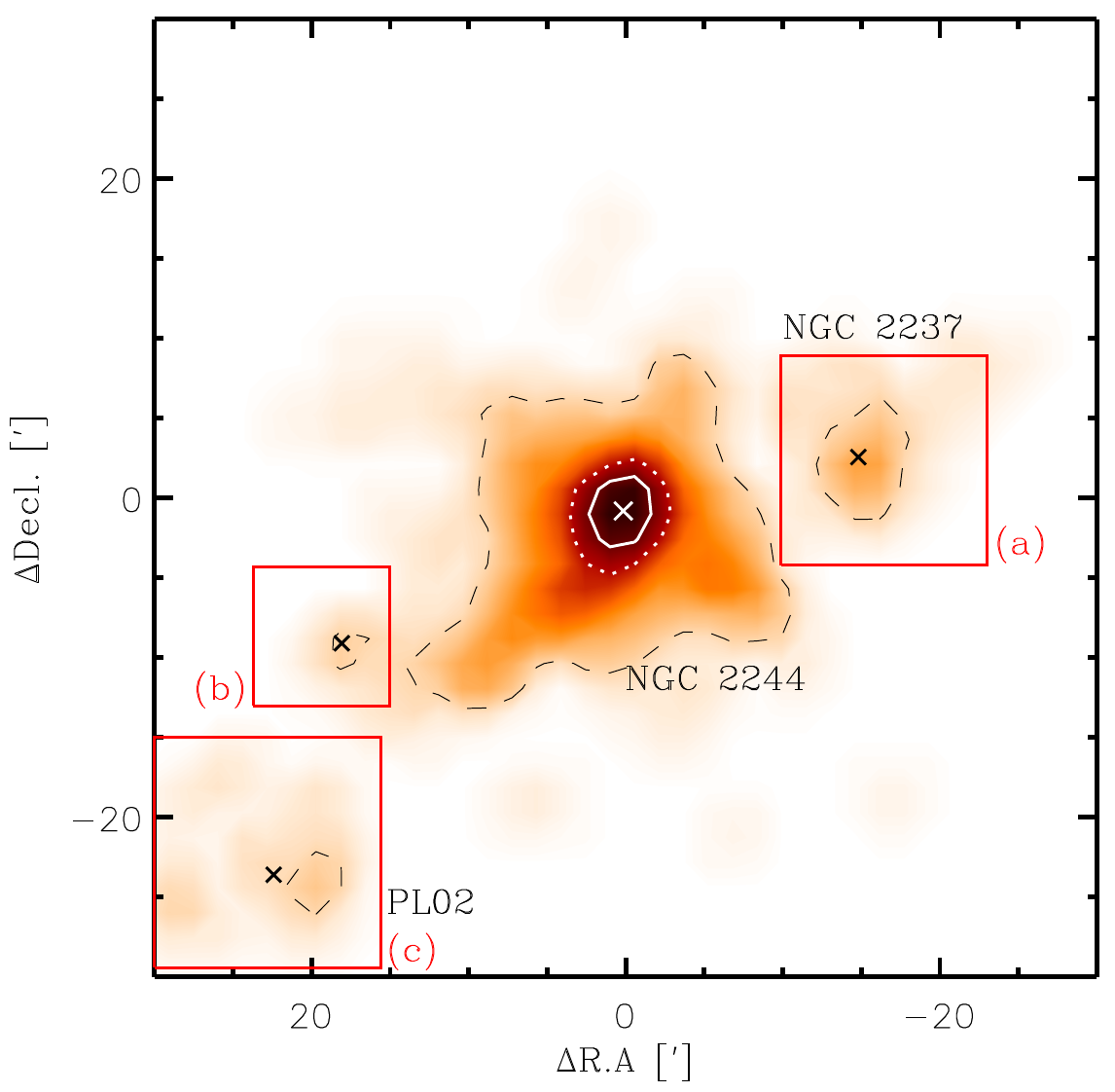}
\caption{Spatial distribution (left) and surface density map (right) 
of members. The background image in the left panel 
was taken from the Digital Sky Survey image. The size of dots is 
proportional to the brightness of individual stars in the $G_{RP}$ 
band. The PM vectors relative to the systemic motion are also plotted 
by solid line, and the different colors represent the position angle 
of the PM vectors as shown in the scale bar. The two circles 
(solid and dashed lines) indicate the apparent radius ($r_{\mathrm{cl}}$) 
and half-mass radius ($r_{\mathrm{h}}$), respectively. Red squares 
denote the locations of three subregions. In the right panel, white solid, 
white dotted, and black dashed lines represent contours corresponding to the 
70, 50, and 8\% levels of the maximum density, respectively. The 
squares correspond to the three subregions in the left panel. The crosses 
represent the centers of the four groups determined from median coordinates. 
The reference coordinate is the same as Figure~\ref{fig1}.}\label{fig4}
\end{figure*}

Figure~\ref{fig3} shows the CMD of the selected members. A number 
of faint stars with large errors on parallax and PMs are naturally 
excluded through our member selection processes. The membership 
of these stars could be improved from the later release of the Gaia 
data. There is an X-ray source that was classified as a Class II 
YSO (the open circle in the upper panel of Figure~\ref{fig3}). This object 
has a lower luminosity than those of the other members at a given color. 
An intermediate-mass YSO with a nearly edge-on disk could be a 
possible explanation for the low luminosity. However, such an object 
rarely shows X-ray emission and has low stellar mass \citep{SBC08,
SSB09}. We therefore cautiously excluded this object from our member list. A total 
of 403 members were finally selected. We present the list of the members 
in Table~\ref{tab2}.

We compared our member list with that of \citet{C-G18}. 
There are 238 stars in common with membership probabilities 
higher than 0.5. However, we suspect that their list contains 
a number of field interlopers below the locus of YSOs in CMD.  
They may be too old to be the members of this young SFR. Some giant stars 
were also selected as members. In addition, the member 
list is missing several O- and B-type stars that are the 
most likely members. For these reasons, we decided 
to use our own list of members.

We present the distance and RV distributions of the 
selected members in the lower panel of Figure~\ref{fig3}. 
The distances of individual stars were computed by the inversion 
of the zero-point-corrected Gaia parallaxes \citep{gedr3,LBB20}.
Since these two distributions appear as Gaussian distributions, 
the distance and systemic velocity of this SFR are obtained 
from the center of the best-fit Gaussian distributions. The 
distance is determined to be $1.4\pm0.1$ (s.d.) kpc. This 
is in good agreement with those derived by previous studies 
within errors \citep{C-G18,HPV00,KHS19,OI81}. 
The systemic RV of this SFR is $13.4\pm2.7$ (s.d.) km s$^{-1}$. 
To minimize the contribution of binaries, we limited the RV sample 
to stars with RVs within three times the standard deviation 
from the peak value.

\section{Structures and kinematics} \label{sec:sec4}
Figure~\ref{fig4} displays the spatial distribution and surface 
density map of the members. The surface density map 
was obtained by counting stars with an interval of $4\farcm0$ 
along the R.A. and Decl., where grids dithered by half the 
bin size were used to improve the spatial resolution. The 
central cluster NGC 2244 seems to have a substructure as 
its surface density does not sharply drop toward 
the southeast. In addition to the cluster, there are at least 
three stellar groups with slightly enhanced peak surface densities 
(8\% of the maximum density of NGC 2244) compared to nearby regions. 
The presence of the group (NGC 2237) at the western patch  
of the Rosette Nebula was reported by previous studies \citep{L05,BB09,
WFT10}. The small eastern group is found at the tip of the gas 
pillar on the east side of the nebula, and the other group corresponding 
to the known PL02 \citep{PL97} is located at the southeastern 
molecular ridge.

We determined the central position and systemic motion of 
NGC 2244 in order to probe stellar motions relative to this 
cluster. The reference coordinate (R.A. = 06$^{\mathrm{h}}$ 
31$^{\mathrm{m}}$ $55\fs00$, Decl. = $+04^{\circ}$ 56$^{\prime}$ 
$30\farcs0$, J2000) used in Figure~\ref{fig1} was adopted 
as the initial position of the cluster center. A new central position 
and systemic PM of NGC 2244 were obtained from the median 
coordinates and PMs of stars within a radius of 5$^{\prime}$. 
This procedure was repeated until these parameters converged 
to constant values. We found the cluster center at R.A. 
$= 06^{\mathrm{h}} 31^{\mathrm{m}} 55\fs60$, Decl. = $+04^{\circ} 55^{\prime} 
41\farcs7$ (J2000) and the systemic PM of $\mu_{\alpha}\cos\delta = -1.731$ 
mas yr$^{-1}$, $\mu_{\delta} = 0.312$ mas yr$^{-1}$. 

\subsection{NGC 2244}
\begin{figure}[t]
\epsscale{1.0}
\plotone{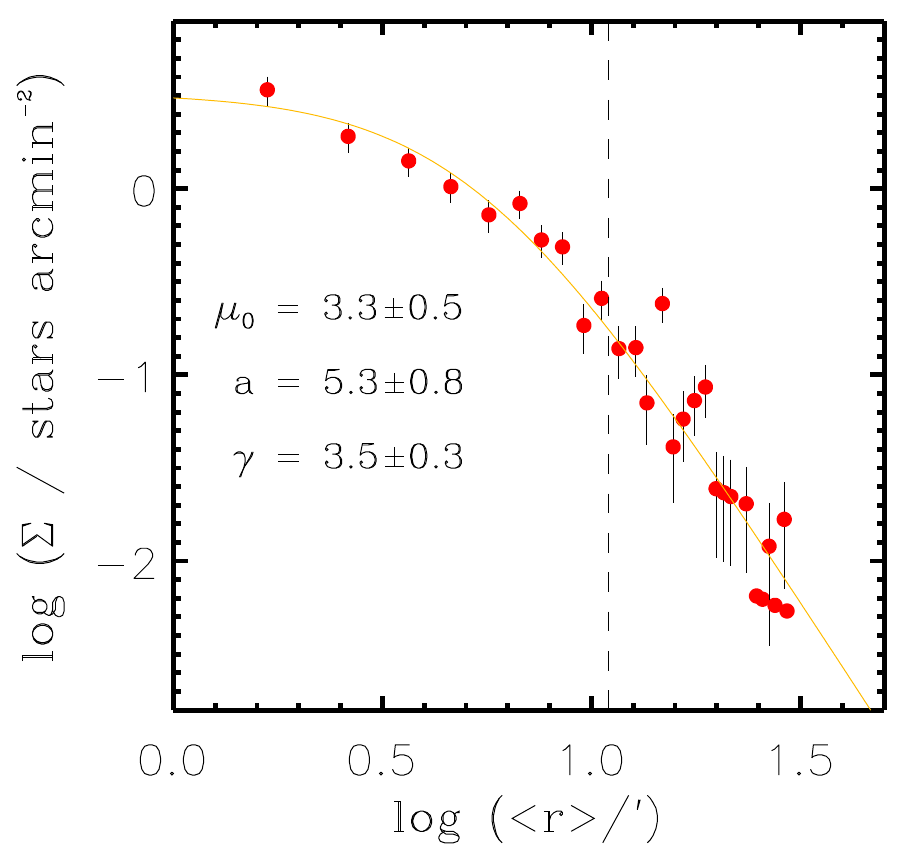}
\caption{Radial surface density profile in logarithmic scale. 
The orange curve represents the best-fit model from 
\citet{EFF87}. The radius at which the western group begins 
to affect the surface density profile is indicated by a dashed 
line ($r_{\mathrm{cl}} \sim 11^{\prime}$). }\label{fig5}
\end{figure}

Figure~\ref{fig5} displays the radial surface density profile 
fit to the model profiles of \citet{EFF87} in a logarithmic scale. 
The surface density of NGC 2244 decreases 
with the radial distance, but is influenced by NGC 2237 
beyond the radius of 11$^{\prime}$ (see also 
the left panel of Figure~\ref{fig4}). Therefore, we adopted this 
boundary as the apparent radius of NGC 2244 ($r_{\mathrm{cl}}$). 
This radius encompasses the majority of cluster members. The 
fundamental parameters and kinematics of this cluster 
were investigated within this radius.

\subsubsection{Age}
In order to derive the fundamental parameters (reddening, 
age, mass, and half-mass radius) of this cluster, we fit the 
isochrones from MESA models considering the effects of 
stellar rotation \citep{CDC16,D16} to the CMD. The distance modulus of 10.73 mag 
(equivalent to 1.4 kpc) was applied to the unreddened 
isochrones ($A_V = 0$ mag). For the early-type members 
with colors bluer than $G_{BP} - G_{RP}$ = 0.6, the reddening 
of the individual stars was obtained by fitting them to the 
distance-corrected isochrones along the reddening 
vector of \citet{WC19} [$A_{\mathrm{GRP}}/E(G_{BP}-G_{RP}) 
= 1.429$]. The mean reddening is then estimated to be 
$\langle E(G_{BP} - G_{RP}) \rangle = 0.733 \pm 0.078$. 

\begin{figure}[t]
\epsscale{1.0}
\plotone{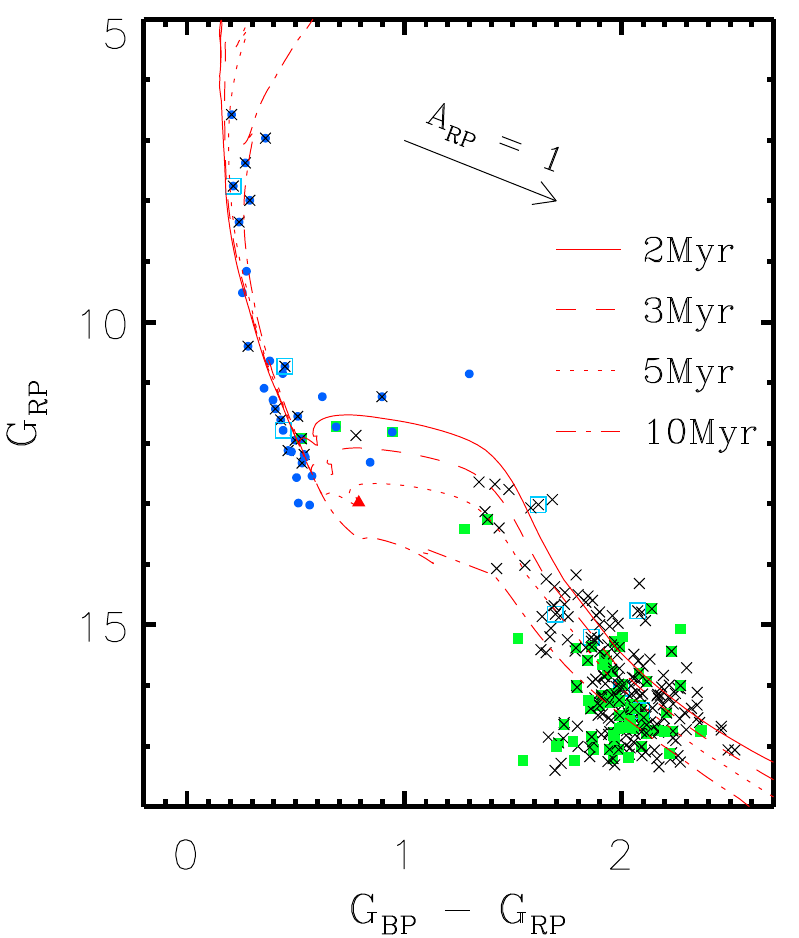}
\caption{CMD of NGC 2244. Red curves are the isochrones 
from the MESA models \citep{CDC16,D16} for 2, 3, 5, and 10 Myrs. 
The arrow represents the reddening vector. The other symbols are 
the same as Figure~\ref{fig1}.} \label{fig6}
\end{figure}

Figure~\ref{fig6} shows the CMD of the members in 
NGC 2244. The isochrones for four different ages (2, 3, 5, 
and 10 Myr) were superimposed on the CMD after correction 
for the mean reddening. Some B-type stars may be already 
evolving into the main-sequence. The magnitude 
of the main-sequence turn-on is sensitive to the age 
of the cluster. The isochrone for 2 Myr well matches 
the main-sequence turn-on. We adopt 2 Myr as the 
age of NGC 2244, and this is in good agreement with 
the previous estimates \citep{HPV00,PS02}. 
However, the entire shape of the CMD seems to be fit 
by isochrones in a wide range of ages ($\Delta$Age $>$ 
7 Myr). There could be systematic uncertainties in the 
calibration adopted in the theoretical models \citep{CDC16,D16}. 
Observationally, photometric errors, imperfect reddening 
correction, binaries, and variabilities of YSOs are 
possible sources of the observed scatter in the CMD 
\citep{LSB15}. In addition, the luminosity spread can 
be caused by multiple star formation events. 

\subsubsection{The initial mass function and total stellar mass}
The mass function is important to estimate the total cluster mass, 
which allows us to infer the virial state of NGC 2244 at a current epoch. To 
derive the mass function of this cluster, we first obtained the masses of individual stars 
from the CMD by means of the isochrones from the MESA evolutionary 
models \citep{CDC16,D16}. The masses of early-type main-sequence 
members ($G_{BP} - G_{RP} <$ 0.6) were obtained by 
interpolating their magnitude to the mass-luminosity relation constructed 
from the isochrones for 2, 5, and 10 Myr, where the individual reddening 
values were applied to the isochrones. For the other members, we adopted 
a mean reddening because the individual reddening values of these stars 
cannot be obtained from the current data. The colors of stars were 
interpolated to the grid of the isochrones with different ages (0.1 -- 
40 Myr) to construct the mass-luminosity relations at the given 
colors. The masses of individual members were then obtained from 
the interpolation of their magnitude to the mass-luminosity 
relations.

To compute the uncertainty in mass estimation, the 
photometric errors were added to or taken out from the observed 
magnitudes and colors. We then found the masses associated to 
these increased or decreased magnitudes and colors in the same way 
as done before. The mass uncertainties were adopted 
from half the difference between the upper and lower values. 
A typical uncertainty due to photometric errors is about 0.1$M_{\sun}$. 
Differential reddening can also affect the mass estimation. The uncertainty 
in mass is then about 0.3 $M_{\sun}$, on average, if we considered the 
differential reddening of 0.078.

We derived the present-day mass function of this cluster by 
counting the number of stars in given logarithmic 
mass bins ($\Delta \log m$). A bin size of 0.2 was used 
to count the number of stars with mass smaller than 
10 $M_{\sun}$, while a larger bin of 1.08 was used 
for higher mass stars because of their paucity. Then, 
the mass function was normalized by the associated 
bin sizes. To avoid the binning effects, we repeated the 
same procedure by shifting by half the bin size. The Poisson 
noise was adopted as the errors on the mass function. 

\begin{figure}[t]
\epsscale{1.0}
\plotone{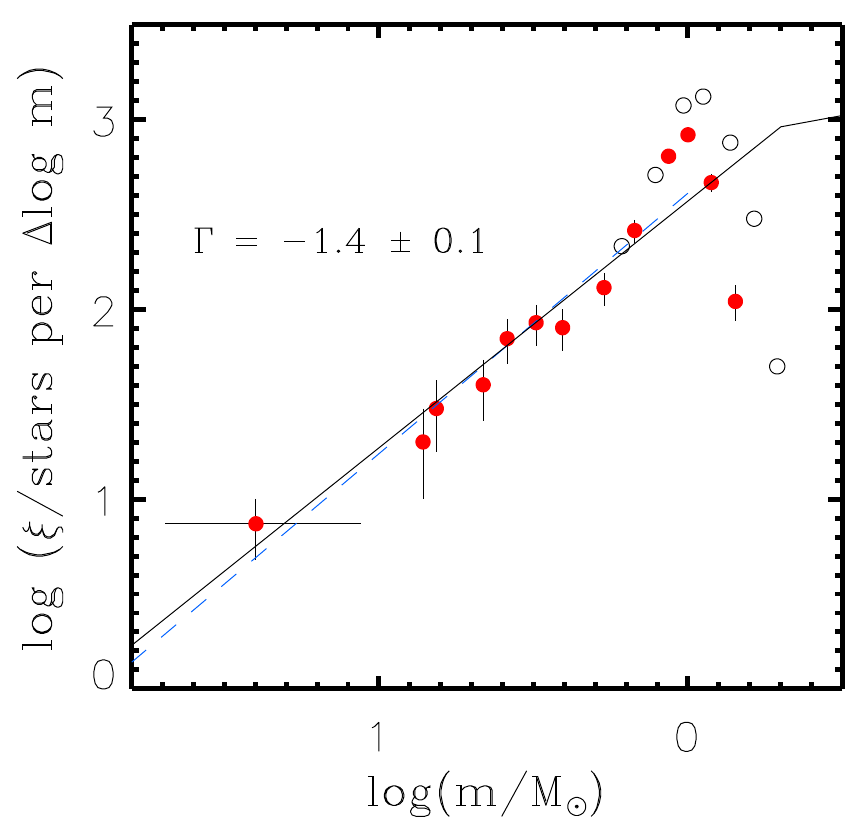}
\caption{Present-day mass function of NGC 2244 (red dots). The result 
of a least square fitting to the mass function is shown by a 
blue dashed line. Black solid lines represent the Kroupa initial 
mass function \citep{K01} for comparison. The open circles denote 
the upper limits of the mass function derived from members and 
member candidates fainter than 14 mag in $G_{\mathrm{RP}}$ band. }\label{fig7}
\end{figure}

Figure~\ref{fig7} displays the present-day mass 
function. The difference between the initial and present stellar 
masses is negligible in the first few Myr. According to the MESA model 
\citep{CDC16,D16}, the most massive star in NGC 2244 has lost 
4$M_\sun$ for 2 Myr. Hence, the present-day mass function can be 
considered as the initial mass function. We constrained the slope ($\Gamma$) 
of the mass function down to the turn-over mass ($\sim$ 1 $M_{\sun}$). 
The slope is estimated to be $\Gamma = -1.4 \pm 0.1$ 
from a least-square fitting method for stars with masses larger than 1 $M_{\sun}$. 
This is in good agreement with those obtained from other SFRs in the Solar 
neighborhood \citep{S55,K01}.

However, it does not guarantee that our member selection is complete 
down to the turn-over mass. Since we selected the most likely members 
using strict selection criteria, many lower-mass members with large 
errors in parallax and PM may have been excluded. We estimated the 
upper limits of the mass function including member candidates fainter 
than 14 mag in $G_{\mathrm{RP}}$ band in the same way as above 
(open circles in Figure~\ref{fig7}). The upper limits follow a trend 
similar to the original one for masses larger than 1 $M_{\sun}$ and 
become larger for the lower mass regime. The number of stars 
obtained from the original mass function is about 30\% lower than 
that from the upper limits at around the turn-over mass. But, this 
may not be the actual completeness limit of our sample because 
some field interlopers may be included in the sample of member 
candidates. If we adopt the upper limits of mass function, then the 
slope of the mass function is estimated to be $\Gamma = -1.6 \pm 0.2$, 
which is consistent with the value found above within errors. 

The sum of stellar masses yields a total mass of 533 
$M_{\sun}$. However, this result does not consider a number 
of stars with very low mass ($< 1 M_{\sun}$). In order to infer 
their population, we scaled the Kroupa initial mass function 
to ours (the original mass function) using the mean difference between these mass 
functions. This scaled Kroupa initial mass function was then 
integrated in the mass range of 0.08 $M_{\sun}$ to 1 $M_{\sun}$. 
The number of cluster members down to the hydrogen burning limit is 
expected to be about 1510$^{+522}_{-375}$, where the upper and lower 
limits were inferred from the integration of the scaled Kroupa initial 
mass functions adjusted by the uncertainty of the scaling factors 
(i.e., the standard deviation) in the low-mass regime ($< 1 M_{\sun}$). 
The total stellar mass of NGC 2244 is estimated to be about 
879$^{+136}_{-98}$ $M_{\sun}$, which is slightly larger than 
those previously obtained by other studies (770$M_{\sun}$ 
-- \citealt{P91}; 625$M_{\sun}$ -- \citealt{BB09}). 

We also estimated a projected half-mass 
radius $(r_\mathrm{h})$ of $5\farcm9$ (equivalent to 2.4 pc) 
from our limited sample. A recent 
study derived the initial mass function of stars in a very 
small central region, down to subsolar mass regime \citep{MSR19}. 
The mass function appeared similar to the Kroupa initial mass 
function, which implies that there is no sign of dynamical 
mass segregation. Hence, $r_{\mathrm{h}}$ obtained in this study 
may not be significantly altered even if the radial mass 
distribution of stars in the full mass range is considered. 

\begin{figure}[t]
\epsscale{1.0}
\plotone{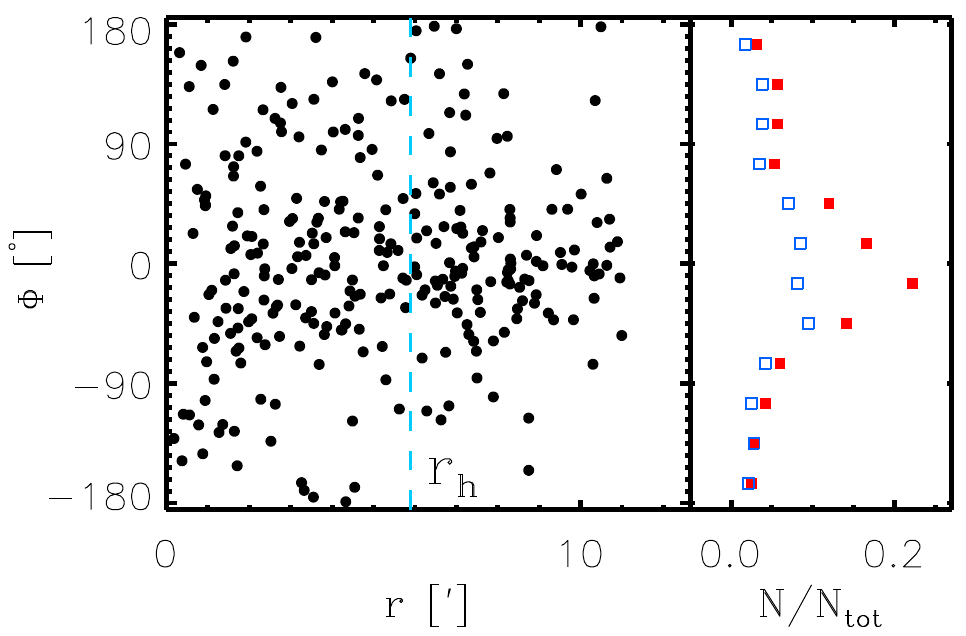}
\caption{$\Phi$ distribution of stars in NGC 2244. The dashed line 
in the left panel represents $r_{\mathrm{h}}$ of this cluster. Red filled 
squares and blue open squares exhibit the histograms of $\Phi$ 
values for all stars and only stars within $r_{\mathrm{h}}$, respectively.}\label{fig8}
\end{figure}

\begin{figure*}[t]
\includegraphics[width=5.5cm]{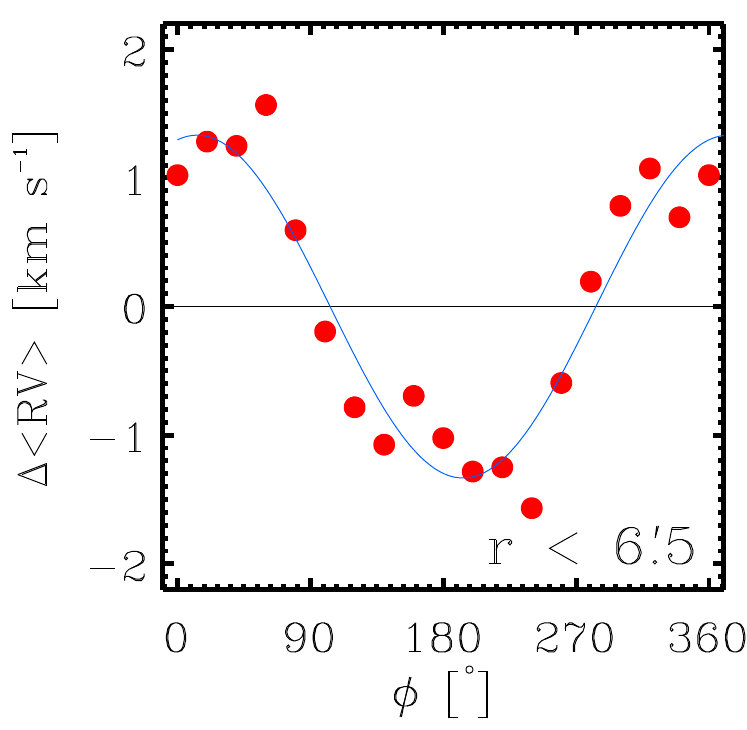}\includegraphics[width=5.5cm]{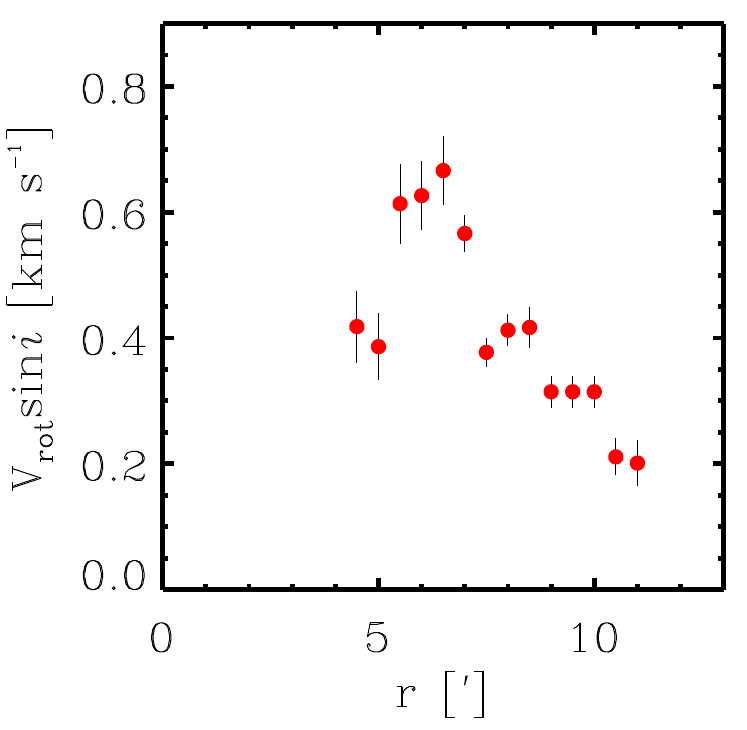}\includegraphics[width=6.8cm]{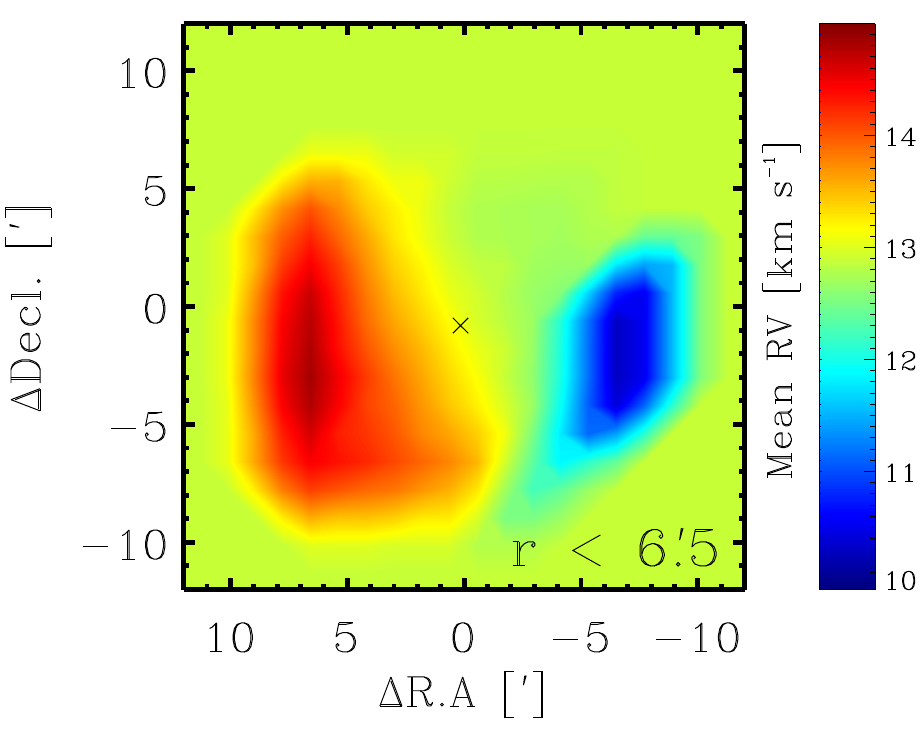}
\caption{Signature of cluster rotation. Left panel: Differences of mean 
RVs with respect to the position angles of the projected rotational axis within 
$r < 6\farcm5$ (2.6 pc). The blue solid line represents the best-fit sinusoidal 
curve. The projected rotational velocity can be estimated from the amplitude 
of the sinusoidal curve divided by 2. Middle panel: Variation 
of projected rotational velocities with respect to the radial distances. The 
dots and error bars were obtained from the best-fit sinusoidal curves for 
different samples. Right panel: Spatial distribution of mean RVs within 
$r < 6\farcm5$. The color bar shows the different levels of mean RVs. 
The mean RVs were computed within spatial bins of $7\farcm0$. A grid 
dithered by the half the bin size was used to sample the stars. The cross 
indicates the center of NGC 2244. The reference coordinate is the same 
as Figure~\ref{fig1}.}\label{fig9}
\end{figure*}

\subsubsection{Internal kinematics}
The PM vectors of the members relative to the systemic 
motion of NGC 2244 are shown in Figure~\ref{fig4}. Stars 
in the central cluster show outward motions, which 
were also detected in \citet{KHS19}. To probe the direction 
of their motion in detail, we calculated the vectorial angle ($\Phi$) 
between the position vectors from the cluster center and 
the relative PM vectors of stars (see also \citealt{LNGR19,LHY20}). 
A $\Phi$ close to 0$^{\circ}$ means that a given star is radially 
moving away from the center of the cluster, while a $\Phi$ of 
180$^{\circ}$ indicates that the star is moving toward the cluster 
center.

Figure~\ref{fig8} displays the $\Phi$ distribution of stars in 
NGC 2244. The cluster members have $\Phi$ values in a wide 
range. We present the histograms of $\Phi$ values in the right 
panel of the figure. There is a peak at 0$^{\circ}$, which indicates 
that some stars are radially escaping from NGC 2244. Given that 
this feature seems weaker within $r_{\mathrm{h}}$ (blue 
open squares in the figure), the majority of these escaping stars may 
be distributed beyond $r_{\mathrm{h}}$.

In order to search for the signature of cluster rotation, the RV 
distribution of stars was investigated as addressed in many 
previous studies \citep{LKL09,MDFY13}. We first considered 
a projected rotational axis across the cluster center from 
north ($0^{\circ}$) to south ($180^{\circ}$). The difference 
between the mean RVs of stars in the two regions separated 
by the projected rotational axis was computed. 
The same procedure was repeated for various position angles 
of the projected rotational axis with an interval of $20^{\circ}$. 

We found a variation of the mean differences of RVs with 
the position angles of the projected rotational axis, as shown 
in the left panel of Figure~\ref{fig9}. This variation was fit to the sinusoidal 
curve as below:
\begin{equation}
\Delta \langle\mathrm{RV}\rangle =  2V_{\mathrm{rot}}\sin i \sin(\phi + \phi_0) 
\end{equation}
\noindent where $V_{\mathrm{rot}}$, $i$, and $\phi_0$ represent the 
rotational velocity, inclination of a rotational axis, and phase, respectively. 
The amplitude of the sinusoidal curve corresponds to twice the 
projected rotational velocity ($V_{\mathrm{rot}}\sin i$). The position angle of 
the projected rotational axis can be inferred from 
$270^{\circ} - \phi_0$. The projected rotational axis is tilted by 
$\sim 20^{\circ} (\pm14^{\circ})$ from north to east. We investigated a variation of 
$V_{\mathrm{rot}}\sin i$ using stars within various radii from 
$4\farcm5$ to $11\farcm0$. The middle panel of Figure~\ref{fig9} 
displays the variation of $V_{\mathrm{rot}}\sin i$ with 
the projected radial distances. In the inner region ($< 6\farcm5$), 
$V_{\mathrm{rot}}\sin i$ of stars increases up to 0.67 km s$^{-1}$ 
with the radial distance, and drops to $\sim$ 0.20 km s$^{-1}$ 
beyond this radius. It is interesting that this cluster rotation is 
evident around $r_{\mathrm{h}}$. A clear pattern of cluster rotation 
is also seen in the position-velocity diagram (the right panel of 
Figure~\ref{fig9}). 

The signature of cluster rotation was also searched for using 
PMs. The one-dimensional tangential velocities ($V_{\mathrm{t}}$) of 
stars along R.A. and Decl. were calculated by adopting a distance 
of 1.4 kpc. Since the diameter of the cluster is very small 
compared to the distance, the errors on $V_{\mathrm{t}}$ by the extent 
of the cluster is less than 1\% under the assumption that 
this cluster has a spherical shape. The rotational velocity on 
the celestial sphere is given by the equation below:
\begin{equation}
V_{\mathrm{rot}} = (d_{\mathrm{R.A.}} \times V_{\mathrm{Decl.}} - d_{\mathrm{Decl.}} \times V_{\mathrm{R.A.}})/r
\end{equation}
\noindent where $r$, $d_{\mathrm{R.A.}}$, $V_{\mathrm{R.A}}$, 
$d_{\mathrm{Decl.}}$, and $V_{\mathrm{Decl.}}$ represent the 
radial distance, the position and $V_{\mathrm{t}}$ in R.A., and 
the position and $V_{\mathrm{t}}$ in Decl., respectively. The 
non-zero mean velocity would yield the rotational velocity of this cluster. 
However, the mean velocity was calculated to be $0.03 \pm 1.32$ 
km s$^{-1}$, meaning that this cluster does not have 
significant azimuthal motion. This result is consistent with that 
found by \citet{KHS19}. Since the cluster rotation was only detected 
in RV, the rotational axis of NGC 2244 may be tilted by 
almost 90$^{\circ}$ from the line of sight.

\subsubsection{Dynamical state}
We computed the virial velocity dispersion of NGC 2244 by using 
the equation \citep{PW16} below : 
\begin{equation}\label{eq:eq4}
\sigma_{\mathrm{vir}} = \sqrt{{2GM_{\mathrm{total}} \over \eta R}}
\end{equation}
\noindent where $G$, $M_{\mathrm{total}}$, $R$, and $\eta$ 
represent the gravitational constant, enclosed total mass, radius, 
and the structure parameter, respectively. Note that we can 
consider the total cluster mass of 879 $M_{\sun}$ as the enclosed 
mass given that gas has already been removed around NGC 2244 
according to $^{12}$CO ($J= 1-0$) and $^{13}$CO ($J=1-0$) 
observations \citep{HWB06}. The $r_{\mathrm{h}}$ of 2.4 pc was used in Equation~\ref{eq:eq4}. 
$\eta$ is a structure parameter with a value between 1 and 12 
depending on the geometry of the cluster. Here, $\eta$ of 10 
was adopted because the observed surface density profile 
($\gamma = 3.5\pm0.3$; Figure~\ref{fig5}) is close to that 
($\gamma = 4$) of \citet{P11} (see also Figure 4 of \citealt{PMG10}). 
The expected velocity dispersion at a virial state is about 
0.6 km s$^{-1}$ (c.f., the virial velocity dispersion for 
other clusters is in the range of 0.4 -- 1.6 km s$^{-1}$; 
\citealt{KHS19}). The virial velocity dispersion is not very 
sensitively varied with the total mass because it is proportional 
to $\sqrt{M_{\mathrm{total}}}$. The virial velocity dispersion 
is almost constant even if we adopted the upper limit of the 
total cluster mass.

\begin{figure}[t]
\epsscale{1.0}
\plotone{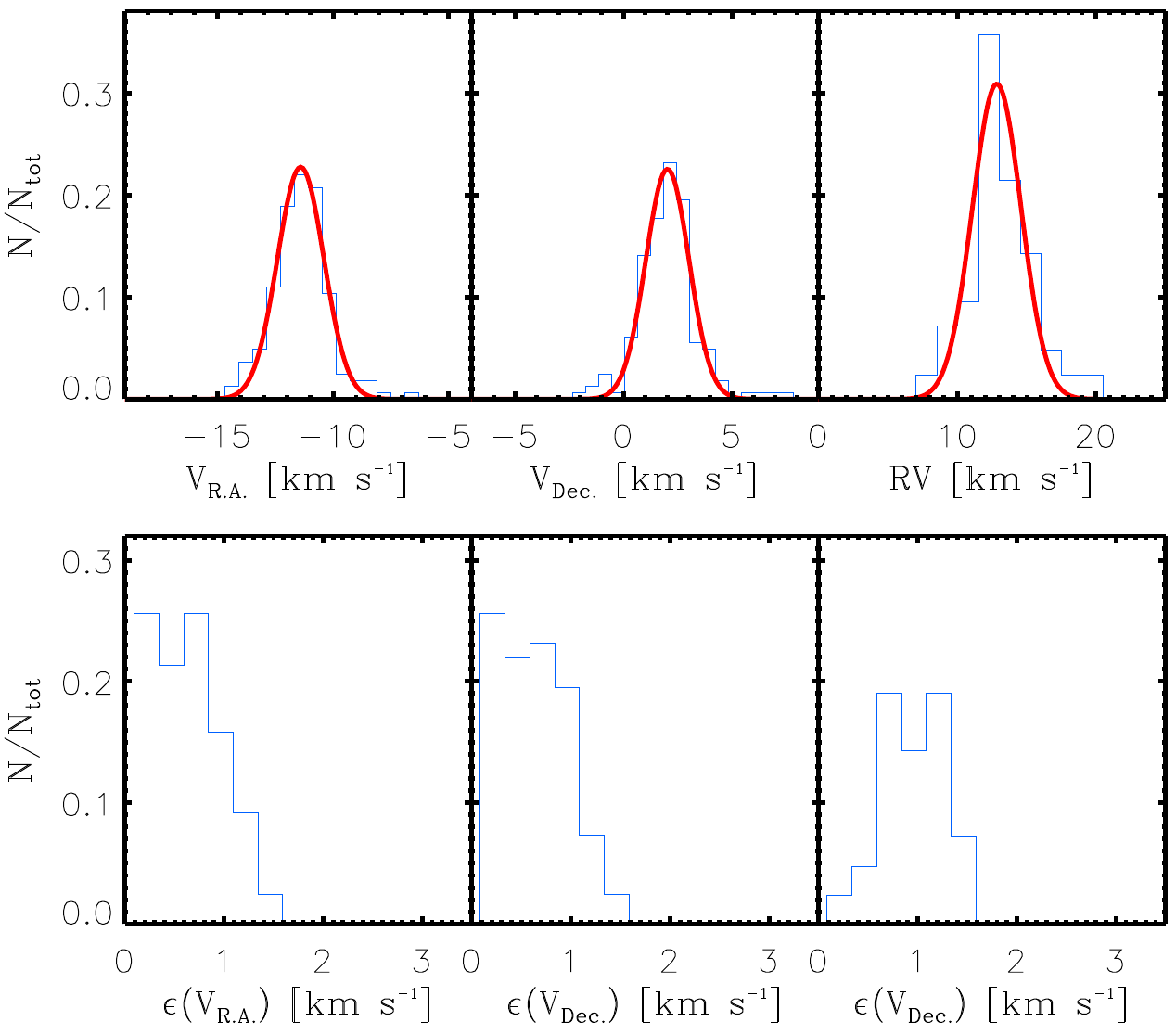}
\caption{Velocity (upper) and error (lower) distributions along R.A., 
Decl., and the line of sight. The histograms were obtained from stars 
within $r_{\mathrm{h}}$ using bin sizes of 0.6, 0.6, and 1.5 km 
s$^{-1}$ for $V_{\mathrm{R.A}}$, $V_{\mathrm{Decl.}}$, and 
RV, respectively. The red curves represent the best-fit 
Gaussian distributions. The systemic velocities and velocity 
dispersions are presented in Table~\ref{tab3}. In the lower panels, 
the associated errors were sampled with a bin size of 
0.25 km s$^{-1}$.}\label{fig10}
\end{figure}

Figure~\ref{fig10} displays the distributions of $V_{\mathrm{R.A}}$, 
$V_{\mathrm{Decl.}}$, and RV (the upper panels) with their error 
distributions (the lower panels). We derived the systemic velocities 
and velocity dispersions from the best-fit Gaussian distributions 
(Table~\ref{tab3}). The typical errors on the measurements are 
estimated from the weighted-mean errors, where the associated 
error distributions were used as the weight values for given errors. 
We obtained the intrinsic velocity dispersions of 0.8 and 0.8 
km s$^{-1}$ along R.A., and Decl., respectively, from quadratic subtraction 
between the observed velocity dispersions and their typical errors. 
The effect of system rotation on the observed RV dispersion can 
be simulated by using the ideal Gaussian velocity distribution based 
on the underlying rotation with position angles. As a result, 
the observed velocity dispersion can be inflated by about 14\% compared 
to the intrinsic velocity dispersion. Since our measurement of the velocity 
dispersion along the line of sight, after quadratic subtraction by the typical 
error of 1.6 km s$^{-1}$, is about 0.8 km s$^{-1}$, the intrinsic 
RV dispersion would be 0.7 km s$^{-1}$. The mean value from the 
intrinsic velocity dispersions along R.A., Decl., and the line of sight 
is $0.8 \pm 0.1$ km s$^{-1}$.

The initial size of this cluster might have been smaller than the 
current one. In addition, a large amount of the natal cloud mass probably 
remained around the cluster. The virial velocity dispersion 
might have thus been larger than what we derived. While the velocities 
of cluster members at a supervirial state are not significantly 
altered in the first several Myrs \citep{SPA19}, the virial 
velocity dispersion could decrease with the evolution of the cluster 
because of the cluster expansion and gas expulsion. At the current epoch, 
the observed velocity dispersion is comparable to $\sqrt{2}$ 
times the virial velocity dispersion, which implies that the total 
kinetic energy is almost in balance with the total potential energy. This cluster may 
be gravitationally unbound. The weak pattern of expansion within 
$r_{\mathrm{h}}$ supports this argument.

\begin{table*}
\begin{center}
\setlength\tabcolsep{5pt}
\caption{Velocity dispersions. \label{tab3}}
\begin{tabular}{lcccc}
\tableline\tableline\scriptsize
               & V$_{\mathrm{sys}}$ &  $\sigma_{\mathrm{obs}}$ & $\sigma_{\mathrm{err}}$ & $\sigma_{\mathrm{int}}$ \\
& (km s$^{-1}$) & (km s$^{-1}$) & (km s$^{-1}$) & (km s$^{-1}$)\\
\hline
$V_{\mathrm{R.A}}$ & -11.4 &1.0 & 0.6&0.8\\
$V_{\mathrm{Decl.}}$ & 2.0&1.0&0.6&0.8\\
RV &12.8 &1.8 &1.6&0.7\tablenotemark{a} \\
\tableline
\end{tabular}
\tablecomments{Velocity dispersions were obtained for members 
within $r_h$. Columns (2), (3), and (4) represent the systemic velocities, 
the observed velocity dispersions, typical errors, and the intrinsic 
velocity dispersions along R.A, Decl., and the line of sight, respectively. }
\tablenotetext{a}{The contribution of the cluster rotation ($\sim$ 10\%) 
has been corrected for the RV dispersion. }
\end{center}
\end{table*}

\begin{figure}[t]
\epsscale{1.0}
\plotone{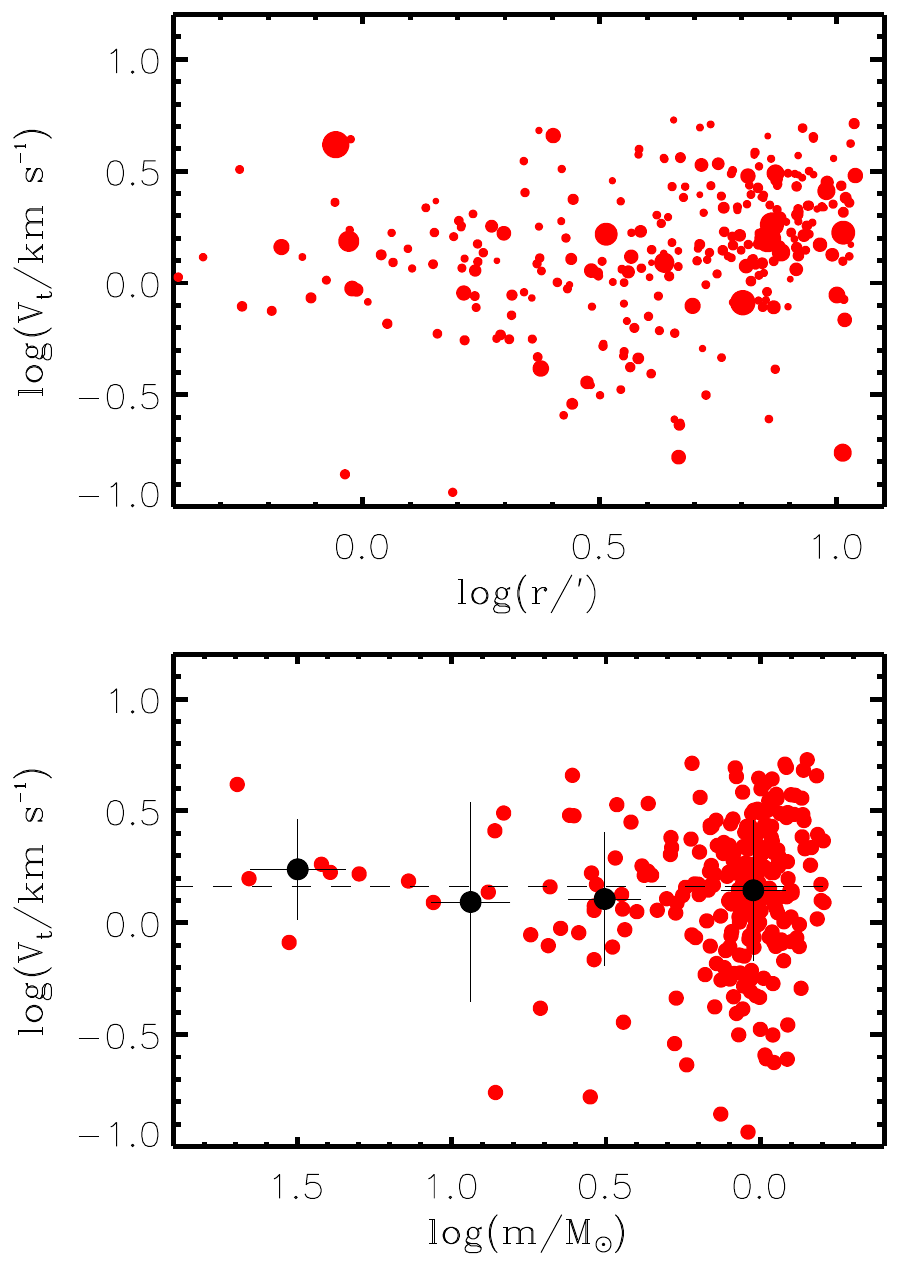}
\caption{Two-dimensional $V_{\mathrm{t}}$ Distribution of cluster 
members with respect to the radial distance (upper) and stellar 
mass (lower). In the upper panel, the size of dots is proportional 
to the mass of individual stars. The dashed line in the lower 
panel represents the median $V_{\mathrm{t}}$. Dots and error 
bars denote the mean and standard deviations of masses and 
$V_{\mathrm{t}}$ within a given logarithmic mass bin of 0.5, 
respectively.}\label{fig11}
\end{figure}

The relaxation time of this cluster was estimated from the 
equation of \citet{BT87}. The crossing time of stars 
for $r_{\mathrm{h}}$ is about 2.9 Myr, adopting the mean 
velocity dispersion of 0.8 km s$^{-1}$. Given the total number 
of stars (1510), the dynamical relaxation time was estimated to be about 76 Myr. 
The age of NGC 2244 is much younger than this timescale, 
and therefore dynamical mass segregation may not be found 
in this cluster. Figure~\ref{fig11} displays the two dimensional 
$V_{\mathrm{t}}$ ($\sqrt{V_{\mathrm{R.A.}}^2 + V_{\mathrm{Decl.}}^2}$) 
distribution with respect to the radial distance and stellar mass. Since 
dynamical mass segregation is the result of energy equipartition 
\citep{BT87}, the signature of this dynamical process can be 
probed from a correlation between stellar mass and velocities, i.e., 
the low-mass stars have velocities higher than those of high-mass 
stars. However, there is no evidence for the energy equipartition 
among cluster members, confirming the result of \citet{CDZ07}. Furthermore, 
the most massive O-type stars are widely distributed across 
the cluster region.  

\subsection{Stellar groups around NGC 2244}
The physical association between stellar groups and 
remaining gas can be a key clue to understanding the 
formation of substructures. To this aim, the distributions 
of stars and gas were investigated in position-velocity 
space. We constructed a three-dimensional 
position-position-velocity diagram of ionized gas as made 
in our previous studies \citep{LSB18,LNGR19}. A data cube 
composed of $90\times90\times90$ regular volume cells 
was created, and then a Delauney triangulation technique 
was used to interpolate the intensities and velocities 
of the forbidden line [N {\scriptsize \textsc{II}}] $\lambda6584$ 
into the individual cells. From this data cube, we also obtained 
position-velocity diagrams by the sum of the counts along R.A. 
and Decl., respectively. The positions and RVs of stars are 
overplotted in these diagrams. 

\begin{figure*}[t]
\includegraphics[width=5.7cm]{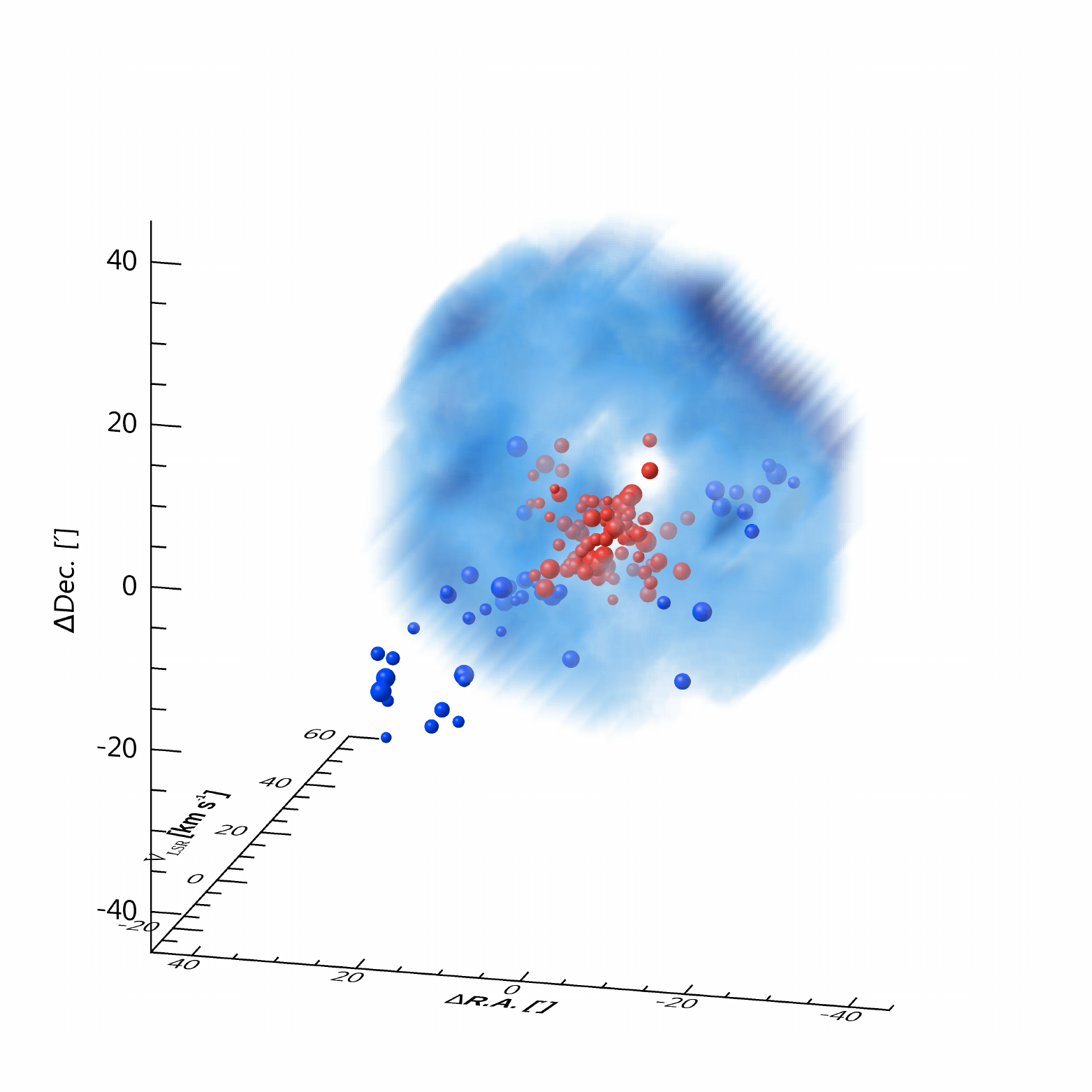}\includegraphics[width=5.7cm]{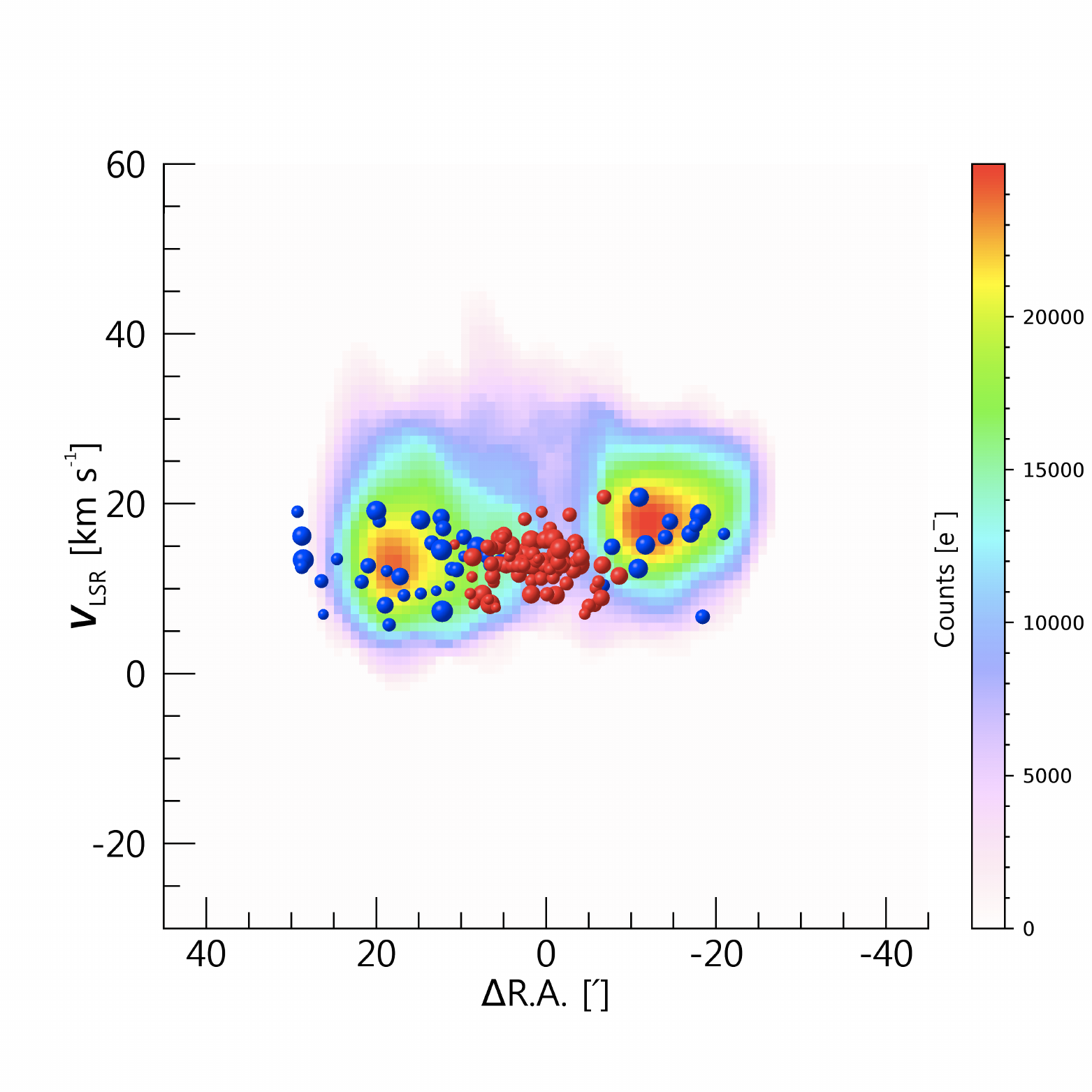}\includegraphics[width=5.7cm]{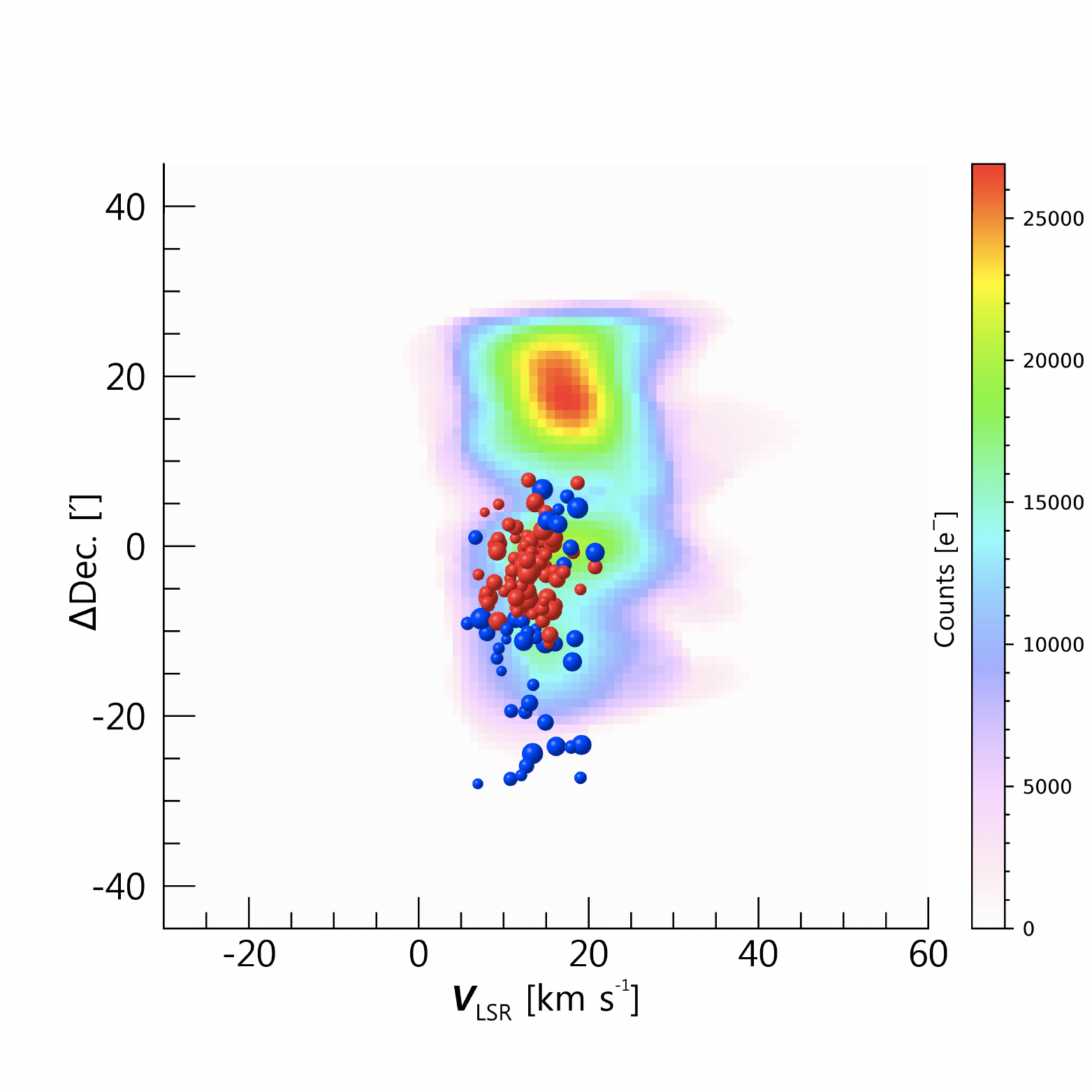}
\caption{Distribution of stars and ionized gas in position-velocity 
diagrams. Stars in the NGC 2244 region are plotted by red spheres, and the 
other stars are shown by blue spheres. The size of the spheres is proportional 
to the brightness of individual stars in the $G_{RP}$ band. Left panel: Position-Position-velocity diagram. 
The bluish nebula represents the distribution of ionized gas traced 
by the forbidden line [N {\scriptsize \textsc{II}}] $\lambda6584$. Middle 
and right panels: Position-velocity diagrams along R.A. and Decl., 
respectively. Contour shows the integrated intensity distribution of 
[N {\scriptsize \textsc{II}}] $\lambda6584$ line in unit of electrons (e$^-$). 
The reference coordinate is the same as Figure~\ref{fig1}. }\label{fig12}
\end{figure*}

\begin{figure*}[t]
\includegraphics[width=6cm]{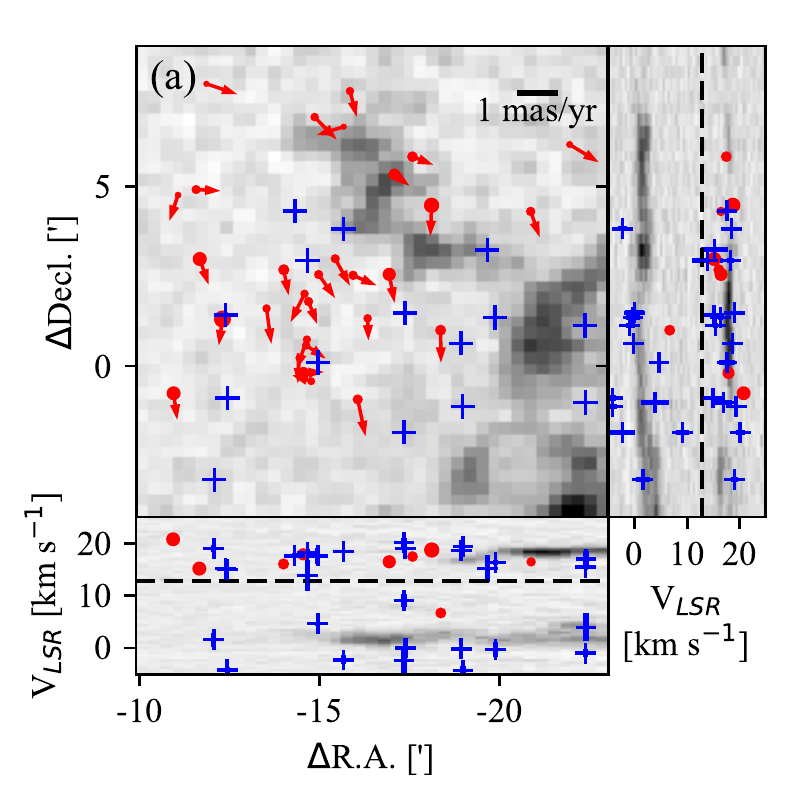}\includegraphics[width=6cm]{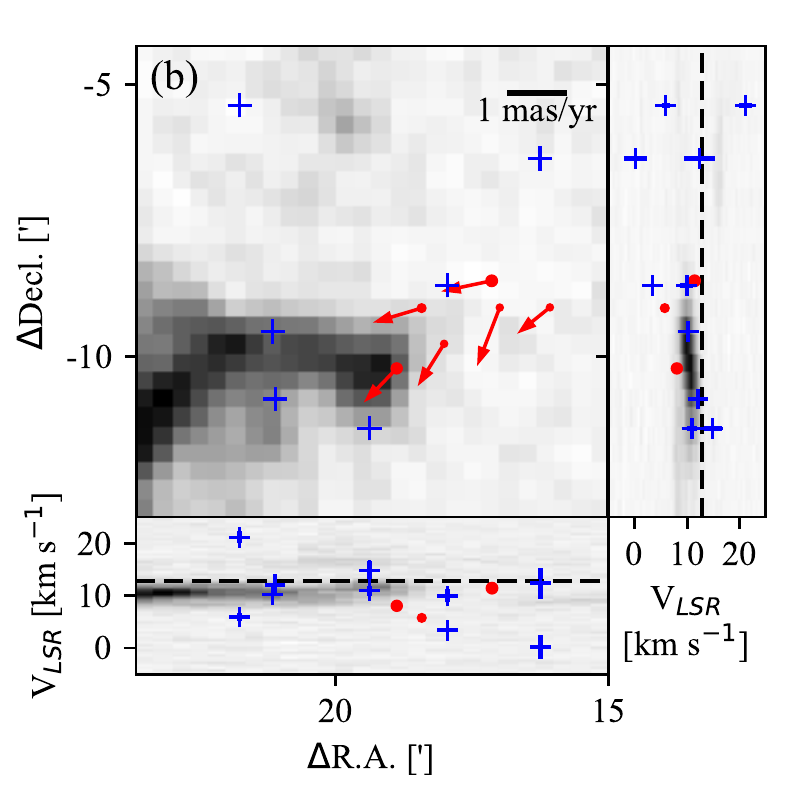}\includegraphics[width=6cm]{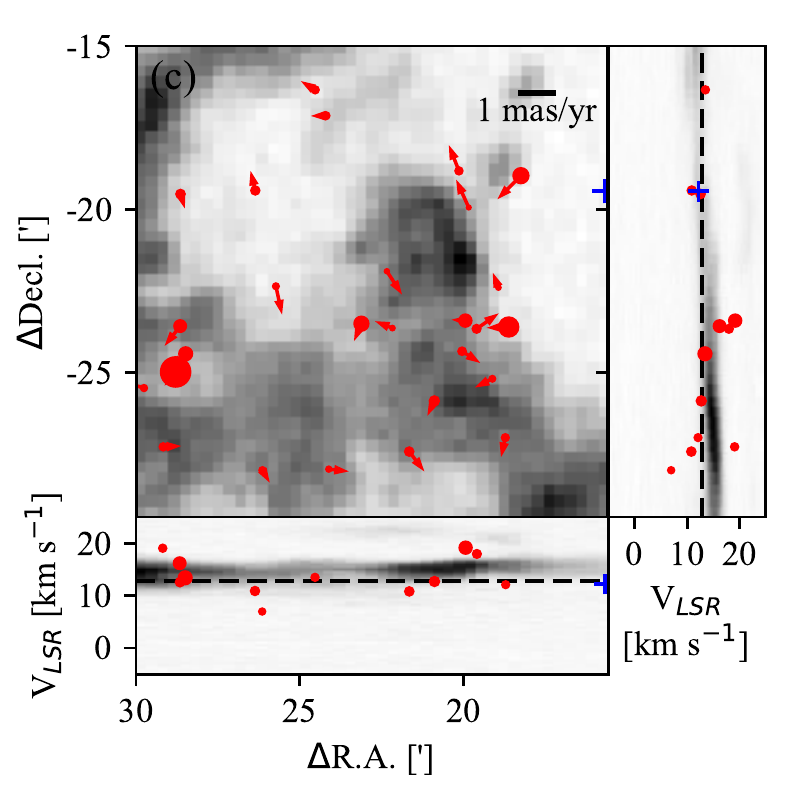}
\caption{Integrated intensity maps and position-velocity diagrams obtained from 
the $^{12}$CO ($J=1-0$) line. The panels {\bf a} to {\bf c} correspond to 
the regions indicated by boxes in the left panel of Figure~\ref{fig4}. The 
distribution of molecular gas is plotted in gray scale. Red dots represent the positions and RVs of 
stars, and their sizes are proportional to their brightness. Arrows denote the PM 
vectors relative to the systemic motion of NGC 2244. Fiber 
positions used to obtain optical spectra as well as RVs of ionized gas are 
shown by pluses. The size of errors on RV is smaller than the symbol size. 
The dashed lines in the position-velocity diagrams represent the systemic 
velocity of this SFR. The reference coordinate is the same as Figure~\ref{fig1}. }\label{fig13}
\end{figure*}

Figure~\ref{fig12} shows the distribution of stars and 
ionized gas in position-velocity space. A cavity of the 
H {\scriptsize \textsc{II}} region is clearly seen in the 
central region occupied by NGC 2244 (red). In the 
position-velocity diagrams, the other stellar groups (blue) 
are found along the H {\scriptsize \textsc{II}} bubble, and they have 
velocities roughly similar to those of the ionized gas. This indicates 
that these groups and remaining gas belong to the same physical system. 
The RVs of ionized gas ranges from 0 km s$^{-1}$ to 30 
km s$^{-1}$, and therefore the Rosette Nebula is expanding 
at velocities up to 17.2 km s$^{-1}$ from NGC 2244 
(12.8 km s$^{-1}$; Table~\ref{tab3}). 

In order to find the physical association between 
stellar groups and gas, it is necessary to probe them 
in detail in smaller position-velocity space. We investigated the three 
stellar groups that are likely associated with adjacent 
gas structures (the left panel of Figure~\ref{fig4}). 
An amount of molecular gas still remains in the vicinity 
of these groups. Figure~\ref{fig13} exhibits 
the distribution of stars on the integrated intensity maps 
and position-velocity diagrams of molecular gas. The RVs of ionized gas 
measured from the [N {\scriptsize \textsc{II}}] 
$\lambda6584$ were also superimposed on the 
figure. The ionized and molecular gas appear to have 
similar RVs.

Most stars in the subregion {\bf a} may be the members of 
the neighboring cluster NGC 2237 \citep{L05,WFT10}. 
There are two gas components in the position-velocity diagrams, 
which is indicative of expanding shell components. The ionized gas 
also traces both components in RV. The heads of the gas pillars 
seen in the left panel of Figure~\ref{fig4} correspond to 
the molecular gas clumps at ($\Delta$R.A., $\Delta$Decl.) 
$\sim$ ($-17^{\prime}$,5$^{\prime}$). The RVs of these clumps 
range 0 km s$^{-1}$ to 5 km s$^{-1}$. Given the systemic velocity of 
NGC 2244 (12.8 km s$^{-1}$), they may be part of the near side of 
the H {\scriptsize \textsc{II}} bubble. Some gas structures 
associated with this bubble extend southwest. This component may be 
expanding at velocities up to $-12.8$ km s$^{-1}$ although this is 
a projected expanding velocity. The other gas component is found 
at ($\Delta$R.A., $\Delta$Decl.) $\sim$ ($-22^{\prime}$,1$^{\prime}$) 
and has RVs in a range of 15 to 20 km s$^{-1}$. This component 
may be part of the bubble receding away from NGC 2244 at projected 
velocities up to 7.2 km s$^{-1}$. NGC 2237 is located in 
the vicinity of the heads of the gas pillars. Most stars in this 
group tend to systematically move toward the south. Stars with RV measurements 
have RVs in range 15 to 21 km, except for one star. Therefore, 
this group seems kinematically associated with the latter gas 
component, rather than the gas pillars.

There is a gas pillar in the subregion {\bf b}. This gas structure 
is moving at velocities of about 10 km s$^{-1}$ along the line of 
sight, which is smaller than the systemic velocity of NGC 2244 (12.8 km 
s$^{-1}$). It implies that this is part of the near side H 
{\scriptsize \textsc{II}} bubble approaching observers. There 
is a small group of stars in the vicinity of the gas pillar. 
These stars have similar RVs to that of the gas pillar although 
the number of stars with RV measurements is only three 
out of six stars. In addition, their PM vectors are systematically 
oriented toward southeast (i.e. away from the central cluster). 

The remaining gas in the subregion {\bf c} is part of 
the northwestern ridge of the Rosette Molecular Cloud. This 
molecular gas has RVs slightly larger than the systemic velocity 
of the central cluster, and therefore it is receding away from 
the cluster in the line of sight. There are more than 20 
members of PL02 in the subregion. Their RVs range from 7 km s$^{-1}$ 
to 19 km s$^{-1}$. It is unclear whether or not PL02 is physically 
associated with this gas due to the spread of the RVs. Given 
that the PM vectors of the PL02 members are small and 
random, the velocity field of this group may be closer to that 
of NGC 2244, than that of the molecular gas. 

\section{Discussion} \label{sec:sec5}
\subsection{The formation of NGC 2244}
The formation of stellar clusters has been explained 
by two scenarios, monolithic formation 
\citep{KAH01,BK13,BK15} and hierarchical mergers 
of subclusters \citep{BBV03,B09}. Cluster rotation 
can also provide additional implication on the cluster 
formation \citep{CLG17,M17}. Here, we discuss the 
formation of the central cluster NGC 2244 in the two-theories 
framework.

Stellar clusters may form along the dense regions in 
a turbulent molecular cloud because star formation 
efficiencies can be high in this environment \citep{K12}. 
Recent simulations of monolithic cluster formation are 
capable of reproducing the internal structures and kinematics of 
clusters \citep{KAH01,BK13,BK15,GBRT17}. Rapid gas 
expulsion and stellar feedback can significantly 
change the structure and kinematics of clusters 
\citep{T78,H80,LMD84,KAH01,BK13,BK15,GBRT17}. However, 
it is also possible to explain the formation of stellar 
clusters even without consideration of these effects. Our previous 
study \citep{LHY20} simulated the dynamical evolution 
of a model cluster formed in an extremely subvirial state. 
This model cluster collapses in the first several million 
years and expands. Subsequently, its structural features 
and kinematics were compared with those of the young open 
cluster IC 1805 (3.5 Myr -- \citealt{SBC17}) composed 
of a virialized core and an expanding halo. The model 
cluster had properties that compared well with the observed 
structure and kinematic properties of IC 1805.

Many other clusters in OB associations also tend 
to show expansion \citep{C-G19,KHS19}, but it seems 
difficult to detect cluster rotation by using only PMs, 
as rotating clusters are rare in the sample of \citet{KHS19}. 
However, the rotation of NGC 2244 was found by 
RVs in this study. Therefore, the spectroscopic observations 
are crucial to perform a more complete search of cluster 
rotation. Indeed, the significant rotation of the young massive 
cluster R136 in the Large Magellanic Cloud was detected from 
spectroscopic observations \citep{HGE12}. The rotational energy accounts 
for more than 20\% of total kinetic energy of this cluster. 
The detection of cluster rotation implies that clusters can form in rotating molecular 
clouds or from hierarchical mergers of small stellar groups with different motions. 

Systematic survey of molecular clouds in 
external galaxies found that molecular clouds are 
rotating \citep{BHR20,REPB03,T11}. More than 50\% of the 
observed molecular clouds have prograde rotation 
with respect to the galactic rotation. The angular 
momenta of these clouds may be acquired from the 
differential rotation of disk and self-gravity. The 
clouds with retrograde rotation can form by 
cloud-cloud collisions \citep{DBP11}. The rotation of 
stellar clusters can thus be inherited from their natal clouds. 
However, it is not yet understood how 
efficiently the angular momenta of their natal clouds 
are redistributed to individual stars and clusters.

The hierarchical merging of gaseous 
and stellar clumps in a turbulent cloud can also lead to the 
formation of rotating clusters \citep{M17}. There are 
some possible candidates. The core of 30 Doradus 
in the Large Magellanic Cloud has an 
elongated shape, and it is composed of two stellar 
populations with different ages \citep{SLG12}. The 
dense group is known to be the young massive cluster 
R136 which presents rotation \citep{HGE12}, while a diffuse stellar 
group is located at the northeast of this cluster. It is 
believed that the elongated structure and age 
difference indicate a recent or an on-going merger of 
these two groups \citep{SLG12}. Similarly, the young 
massive cluster Westerlund 1 has an elliptical shape 
\citep{GBSH11} and hosts two different age groups 
\citep{LCS13}. However, the rotation of this cluster 
has not been reported yet.

NGC 2244 could monolithically form in a rotating 
cloud or filament hub, after which it might undergo cold 
collapse depending on its initial virial state. 
Some stars are now escaping from the cluster as seen in 
Figure~\ref{fig8}. Gas expulsion may also play a role in 
dispersing the cluster members. In addition, there are 
several stars in the north and south beyond $r_{\mathrm{c}l}$, and 
these stars are also moving away from the cluster 
(see Figure~\ref{fig4}). We traced their positions back using 
PMs. As a result, about 85\% (11/13) of these stars were 
in the cluster ($< r_{\mathrm{cl}}$) 2 Myr ago. Therefore, 
the spatial distribution of these stars can also be explained by the 
dynamical evolution of this cluster after the monolithic 
formation.

The observed features in IC 1805 could be explained 
by the monolithic collapse and rebound \citep{LHY20}. In the 
same context, we compare the pattern 
of expansion in NGC 2244 with those in IC 1805. While IC 1805 
shows a strong expansion pattern 
in its outer region \citep{LHY20}, NGC 2244 has, in comparison, 
a rather weak tendency of expansion (Figure~\ref{fig8}). 
One possible explanation would be that the initial 
rotation supported the system against the sub-virial 
collapse. Dynamical mass segregation is a natural 
consequence of the violent relaxation during the cold 
collapse \citep{AGP09}. However, the absence of notable 
mass segregation in NGC  2244 also implies that the cluster 
did not have a strong cold collapse in the past.

The western half of the cluster has a negative 
net RV relative to the systemic velocity as it rotates (Figure~\ref{fig9}), 
while NGC 2237 has a positive 
net RV (Figures~\ref{fig12} and ~\ref{fig13}). The discrepancy 
between the directions of the cluster rotation and 
stellar motions in the subregion cannot be simply explained 
by the monolithic cluster formation in a rotating cloud. 
On the other hand, star formation is still taking place 
along the filaments in the Rosette Molecular Cloud, forming 
several groups of stars \citep{CMF13,PRG08,SMB10}. 
In particular, a clustering of the stellar groups PL04, PL05, 
PL06, and REFL08 is noteworthy \citep{CMF13} (see also 
Cluster E in \citealt{PRG08}). This circumstance is a favorable 
condition for merging of stellar groups. Similarly, 
NGC 2244 has a substructure toward the southeast 
(Figure~\ref{fig4}). The structural feature and rotation 
of this cluster may be the result of hierarchical mergers 
of stellar groups.

\begin{figure*}[t]
\epsscale{1.0}
\plotone{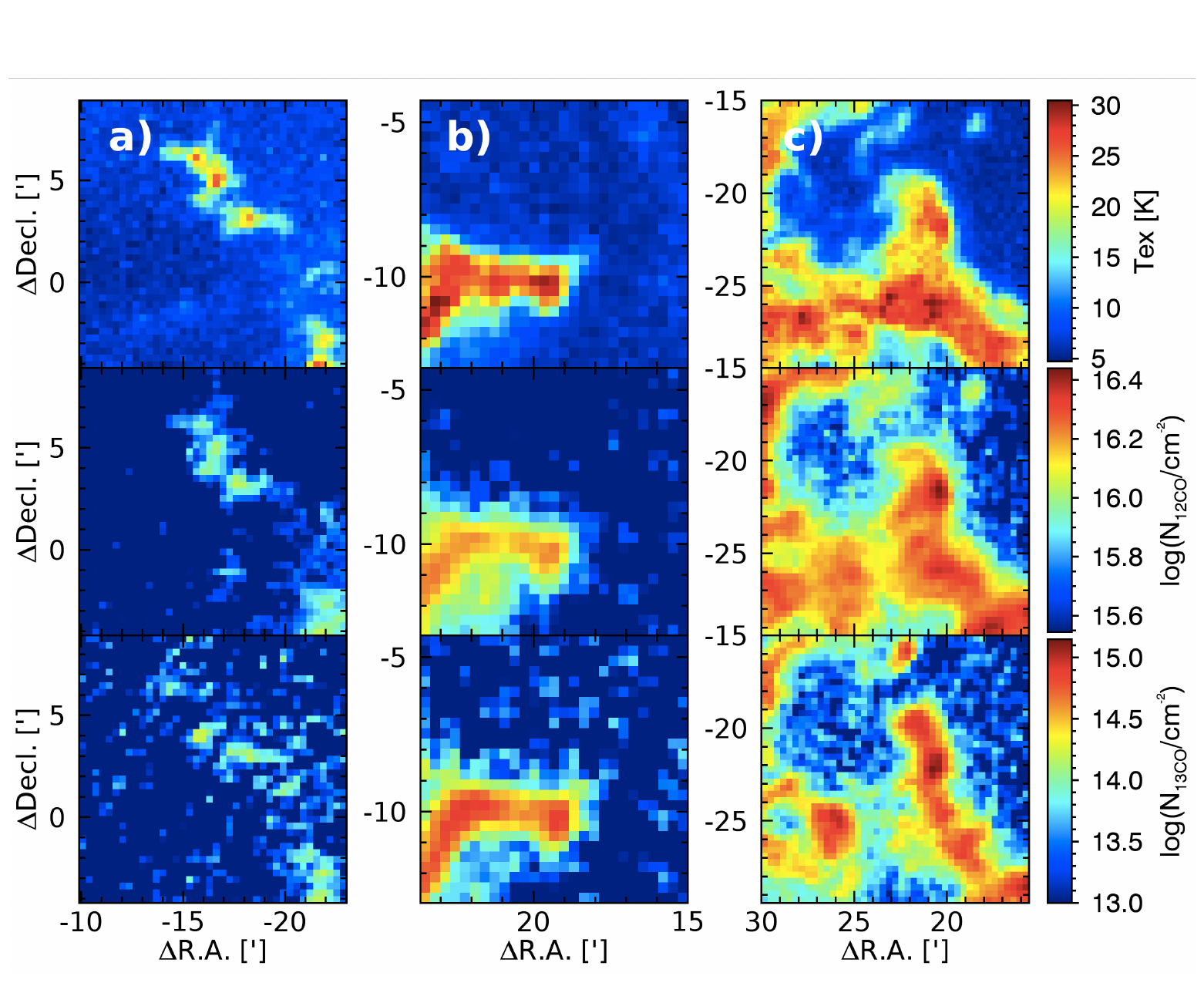}
\caption{Distributions of $T_{\mathrm{ex}}$ and column 
density in the subregions {\bf a} (left), {\bf b} (middle), 
and {\bf c} (right). The levels of 
$T_{\mathrm{ex}}$ and column density are shown by the color bars.  
To minimize the contribution of diffuse gas toward this 
SFR \citep{HWB06}, the velocity ranges were limited to 
$-3.0$ -- 10.0 km s$^{-1}$ for the subregion {\bf a}, 4.0 -- 15.0 
km s$^{-1}$ for {\bf b}, and 6.0 -- 18.5 km s$^{-1}$ for 
{\bf c}. The reference coordinate is the same as 
Figure~\ref{fig1}.}\label{fig14}
\end{figure*}

\subsection{The origin of the stellar population around NGC 2244}
The {\it Herschel} observation reveals that the Rosette Molecular 
Cloud is being affected by the ultraviolet irradiation from the 
massive stars in NGC 2244 \citep{SMB10}. A gradient of dust 
temperature was found toward the southeast. The age differences 
among individual stellar groups indicate the presence of an 
age sequence among YSOs from the H {\scriptsize \textsc{II}} bubble 
toward the Rosette Molecular Cloud. \citet{SMB10} 
argued that the formation of the stellar groups may be 
triggered by the radiative feedback from massive stars in 
NGC 2244. 

However, this argument was later refuted by \citet{CMF13}. 
They measured the slope of spectral energy distributions 
of individual YSOs and better estimated the ages of the stellar 
groups. The stellar groups close to the Rosette Nebula 
are systematically younger than NGC 2244. On the other 
hand, the stellar groups PL06 and PL07 are far from 
the Rosette Nebula, but their ages are comparable to that 
of NGC 2244. As a conclusion, their age distribution could not be fully 
understood in the context of the feedback-driven star 
formation \citep{EL77}. Star formation in the Rosette Molecular 
Cloud seems to take place in two different ways \citep{PL97}. Several 
previous studies also reached similar conclusion on the 
formation of these stellar groups \citep{PRG08,REFL08,YLR13} .

To understand the formation of the three groups that 
we defined at the border of the Rosette Nebula (the left panel 
of Figure~\ref{fig4}), it is necessary to consider both 
models of hierarchical star formation along filaments 
\citep{BSCB11} and feedback-driven star formation 
\citep{EL77} as suggested by \citet{PL97}. Figure~\ref{fig14} 
displays the distributions of $T_{\mathrm{ex}}$ and column 
density along the remaining molecular gas in these subregions. 
The heads of gas pillars in the subregion {\bf a} have $T_{\mathrm{ex}}$ 
and column densities lower than those in the other subregions. 
The column density traced by the optically thin $^{13}$CO 
line compared to $^{12}$CO line shows no agglomeration 
deeper inside the pillar heads. 

A large amount of molecular gas still remains in the 
subregions {\bf b} and {\bf c}. 
$T_{\mathrm{ex}}$ in the subregions 
{\bf b} and {\bf c} appears higher than 25K, which is consistent 
with dust temperature \citep{SMB10}. The molecular 
gas in the subregion {\bf b} condenses along the midplane, 
and a local condensation is found at the ridge of the 
molecular gas in the subregion {\bf c}. These results 
indicate that molecular gas is compressed along the ridge 
of the Rosette Molecular Cloud because the clouds are 
exposed to the ultraviolet irradiation from the O-type stars 
in NGC 2244. Hence, the subregions {\bf b} and {\bf c} 
might be the possible sites of feedback-driven star formation, 
compared to the subregion {\bf a}. If PL02 is physically 
associated with the adjacent molecular gas, its age 
younger than NGC 2244 \citep{CMF13} may support 
this argument.

Here, we discuss the formation of the three groups 
based on their kinematics as shown in Section 4.2.
If stars form in the remaining gas compressed by the 
expanding H {\scriptsize \textsc{II}} bubble, it is expected 
that they are receding away from ionizing sources in 
NGC 2244 and have similar velocities to that of the gas. 
The stars in the subregion {\bf b} seem to follow this 
expectation. These stars are moving away from the 
cluster (toward southeast) and have similar RVs to that 
of the adjacent gas pillar. In addition, this group contains 
a low-mass Class I YSO at an early evolutionary stage 
of protostars. This group may be slightly younger than 
the other groups, and hence its formation might be 
triggered by the feedback from massive stars in NGC 2244. 
We note that there is another Class I object in NGC 2244, 
but this may be a slightly old Herbig Ae/Be star candidate 
with a thick disk (Figure~\ref{fig6}). 

NGC 2237 in the subregion {\bf a} does not show any 
physical association between stars and the gas pillars 
(Figure~\ref{fig13}). Also, this group is moving toward 
south instead of west (the direction of radial expansion 
of the H {\scriptsize \textsc{II}} bubble). This confirms 
that the region {\bf a} is probably not a site of feedback-driven 
star formation. Although the subregion {\bf c} was 
considered as a possible site of feedback-driven 
star formation in \citet{PL97}, the kinematic 
properties of PL02 and adjacent clouds do not support 
the claim. If PL02 was formed by the compression of 
the clouds, the members of PL02 should show PMs systematically 
receding away from NGC 2244. However, they have 
almost random motions, rather than any systematic 
motion toward the southeast as seen in the subgroup 
{\bf b}. In addition, the physical association between 
PL02 and the remaining clouds is unclear. Hence, the 
formation of this group seems not to be related to feedback 
from the massive stars in the cluster. These 
two groups NGC 2237 and PL02 might have been 
independently formed through the hierarchical 
fragmentation of a molecular cloud. 

\section{Summary} \label{sec:sec6}
Stellar associations are not only ideal targets to understand 
star formation process on different spatial scales, but also 
the factories of field stars in the Galactic thin disk. There 
are three theoretical models to explain the formation of 
these young stellar systems, however our observational knowledge 
of their formation is still incomplete. In this study, we investigated gas 
and young stellar population in the Rosette Nebula, the 
most active SFR in the Mon OB2 association. 

We identified the young star members based on photometric 
and spectroscopic criteria, complemented by parallax and proper 
motion criteria based on the recent Gaia astrometric data \citep{gedr3}. A total 
of 403 stars were selected as the members. Their spatial distribution 
showed that this SFR is highly substructured. 

The central cluster NGC 2244 is the most populous group 
in the Rosette Nebula. The age of this cluster was estimated 
to be about 2 Myr by means of stellar evolutionary models 
\citep{CDC16,D16}. We derived its mass function, which 
appeared to be consistent with the Salpeter/Kroupa initial 
mass function \citep{K01} for stars with masses larger than 
1 $M_{\sun}$. However, a number of bona-fide members 
with smaller masses could have been excluded because 
of our strict criteria of member selection. The total number of 
members and cluster mass were deduced to be about 
1510$^{+522}_{-375}$ and 879$^{+136}_{-98}$, respectively. This 
cluster showed a clear pattern of expansion beyond the 
$r_{\mathrm{h}}$ (2.4 pc) and rotation below that radius. 
We also investigated correlations between stellar mass, velocity, 
and the radial distance of stars, but could not find any sign 
of dynamical mass segregation.

Several groups of stars were found at the border of 
the Rosette Nebula. We investigated the kinematic properties 
of stars in three subregions by comparing their RVs with 
those of adjacent gas structures. The eastern group (subregion {\bf b}) 
seemed to be physically associated with the gas pillar and 
receding away from the central cluster. In addition, a low-mass 
Class I object was found in this group. These results can be 
understood in the context of feedback-driven star formation 
\citep{EL77}. On the other hand, there was no kinematic evidence of 
feedback-driven star formation in other subregions. These groups 
might have originated from spontaneous star formation along filaments. 

In conclusion, all the processes proposed by the three theoretical 
models seem to be involved in the formation of this association. 
We need to examine other Galactic OB associations with 
the same perspective as that of this study. A systematic survey 
of both stars and gas in many stellar associations will provide key 
clues to understanding their formation process.

\acknowledgments \label{sec:ack}

The authors thank the anonymous referee for many comments and suggestions. 
The authors would also like to express thanks to Professor Mark Heyer 
for providing supplementary data, Dr. Nelson Caldwell, and ShiAnne 
Kattner for assisting with Hectochelle observations. Observations reported here 
were obtained at the MMT Observatory, a joint facility of the University of Arizona 
and the Smithsonian Institution. In addition, this paper has made use of data 
obtained under the K-GMT Science Program (PID: MMT-2019B-1) 
funded through Korean GMT Project operated by Korea Astronomy 
and Space Science Institute (KASI) and from the European Space Agency (ESA) 
mission {\it Gaia} (https://www.cosmos.esa.int/gaia), processed by the {\it Gaia}
Data Processing and Analysis Consortium 
(DPAC, https://www.cosmos.esa.int/web
/gaia/dpac/consortium). Funding for the DPAC has been 
provided by national institutions, in particular the institutions 
participating in the {\it Gaia} Multilateral Agreement. The Digitized Sky 
Surveys were produced at the Space Telescope Science Institute under 
U.S. Government grant NAG W-2166. The images of these surveys are 
based on photographic data obtained using the Oschin Schmidt Telescope 
on Palomar Mountain and the UK Schmidt Telescope. The plates were 
processed into the present compressed digital form with the permission 
of these institutions. This paper also made use of data products from 
the Wide-field Infrared Survey Explorer, which is a joint project of the 
University of California, Los Angeles, and the Jet Propulsion Laboratory/California 
Institute of Technology, funded by the National Aeronautics and Space Administration. 
IRAF was distributed by the National Optical 
Astronomy Observatory, which was managed by the Association of 
Universities for Research in Astronomy (AURA) under a cooperative 
agreement with the National Science Foundation. This work was supported 
by the National Research Foundation of Korea (NRF) 
grant funded by the Korean government (MSIT) (Grant No: 
NRF-2019R1C1C1005224). Y.N. acknowledges support from the Fonds 
National de la Recherche Scientifique (Belgium), the European Space Agency 
(ESA) and the Belgian Federal Science Policy Office (BELSPO) in the 
framework of the PRODEX Programme (contracts linked to XMM and Gaia). 
J.H. acknowledges support from the Basic Science Research Program 
through the NRF funded by the Ministry of Education 
(No. 2020R1I1A1A01051827). N.H. acknowledges support from the 
Large Optical Telescope Project operated by KASI. B.-G.P. acknowledges 
support from the K-GMT Project operated by KASI. \\

%% To help institutions obtain information on the effectiveness of their 
%% telescopes the AAS Journals has created a group of keywords for telescope 
%% facilities.
%
%% Following the acknowledgments section, use the following syntax and the
%% \facility{} or \facilities{} macros to list the keywords of facilities used 
%% in the research for the paper.  Each keyword is check against the master 
%% list during copy editing.  Individual instruments can be provided in 
%% parentheses, after the keyword, but they are not verified.

\vspace{5mm}
\facilities{MMT (Hectochelle)}

%% Similar to \facility{}, there is the optional \software command to allow 
%% authors a place to specify which programs were used during the creation of 
%% the manuscript. Authors should list each code and include either a
%% citation or url to the code inside ()s when available.

%\software{astropy \citep{2013A&A...558A..33A},  
%          Cloudy \citep{2013RMxAA..49..137F}, 
 %         SExtractor \citep{1996A&AS..117..393B}
 %         }

%% Appendix material should be preceded with a single \appendix command.
%% There should be a \section command for each appendix. Mark appendix
%% subsections with the same markup you use in the main body of the paper.

%% Each Appendix (indicated with \section) will be lettered A, B, C, etc.
%% The equation counter will reset when it encounters the \appendix
%% command and will number appendix equations (A1), (A2), etc. The
%% Figure and Table counter will not reset.

\clearpage
\appendix
\restartappendixnumbering
\section{YSO Classification} \label{sec:appa}
\begin{figure}[t]
\epsscale{1.0}
\plotone{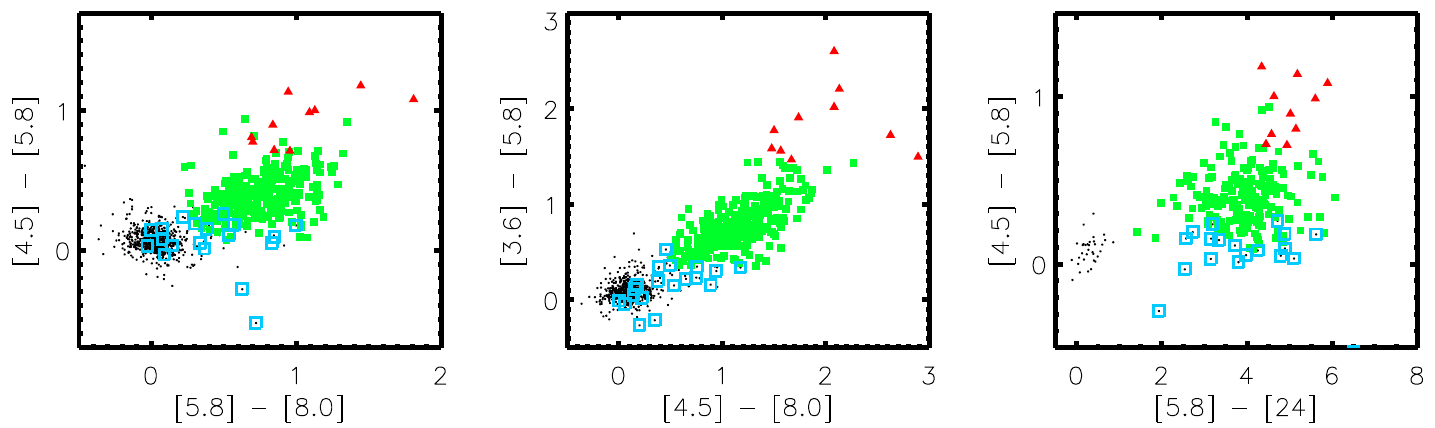}
\caption{Color-color diagrams of stars in the Rosette Nebula obtained 
from \citet{BMR07}. Red triangle, green square, cyan open square, and 
black dot represent Class I, Class II, YSO with a transitional disk, and 
stars without any infrared excess emission, respectively. }\label{afig1}
\end{figure}

The Spitzer IRAC and MIPS 24$\micron$ observations of young stars 
in the Rosette Nebula were carried out by \citet{BMR07}. This survey 
covers an optically visible $30^{\prime}\times30^{\prime}$ region 
in the center of the nebula. We took their photometric data of 1084 
stars detected in all IRAC four bands. Figure~\ref{afig1} displays the 
color-color diagrams of stars in mid-infrared wavelengths. Some stars 
have photospheric colors close to zero, while the others exhibit 
significant mid-infrared excess originating from their warm circumstellar 
disks or envelopes. 

Prior to YSO classification, we flagged the sources of contamination. Active 
galactic nuclei (AGNs) and star-forming galaxies are also sources with 
infrared excess emission \citep{DRPB08}. These objects were identified 
by using the criteria of \citet{GMM08}. A total of 104 AGNs and four 
star-forming galaxy candidates were found. One source was identified 
as a blob of shock emission that appears bright in 4.5 $\micron$ 
\citep{GMM08,SHMP06}. 

The YSO candidates were identified from these color magnitude diagrams 
according to the classification scheme of \citet{GMM08}. We classified 
ten and 262 young stars as Class I and Class II objects, respectively. In 
addition, a total of 19 YSOs with a transitional disk were found in the 
color-color diagram combined with MIPS 24$\micron$ photometry. 
\citet{BMR07} identified 337 Class II and 25 Class I objects from the 
IRAC color-color diagrams and 213 Class II and 20 Class I objects from 
the IRAC and MIPS 24$\micron$ photometry. The results of their YSO 
classification were different from ours for the same data. One reason 
may be due to the adoption of different classification schemes for AGNs, 
star-forming galaxies, and YSOs. The second possible reason is that 
\citet{BMR07} published only the good quality photometric data of sources detected 
in the four IRAC bands, and so it is possible that faint sources were not 
be contained in their data. It is worth noting that such faint YSOs would not 
be selected as final members because of their low-quality astrometric 
data (see Section~\ref{sec:sec3}). 

\begin{figure}[t]
\epsscale{0.7}
\plotone{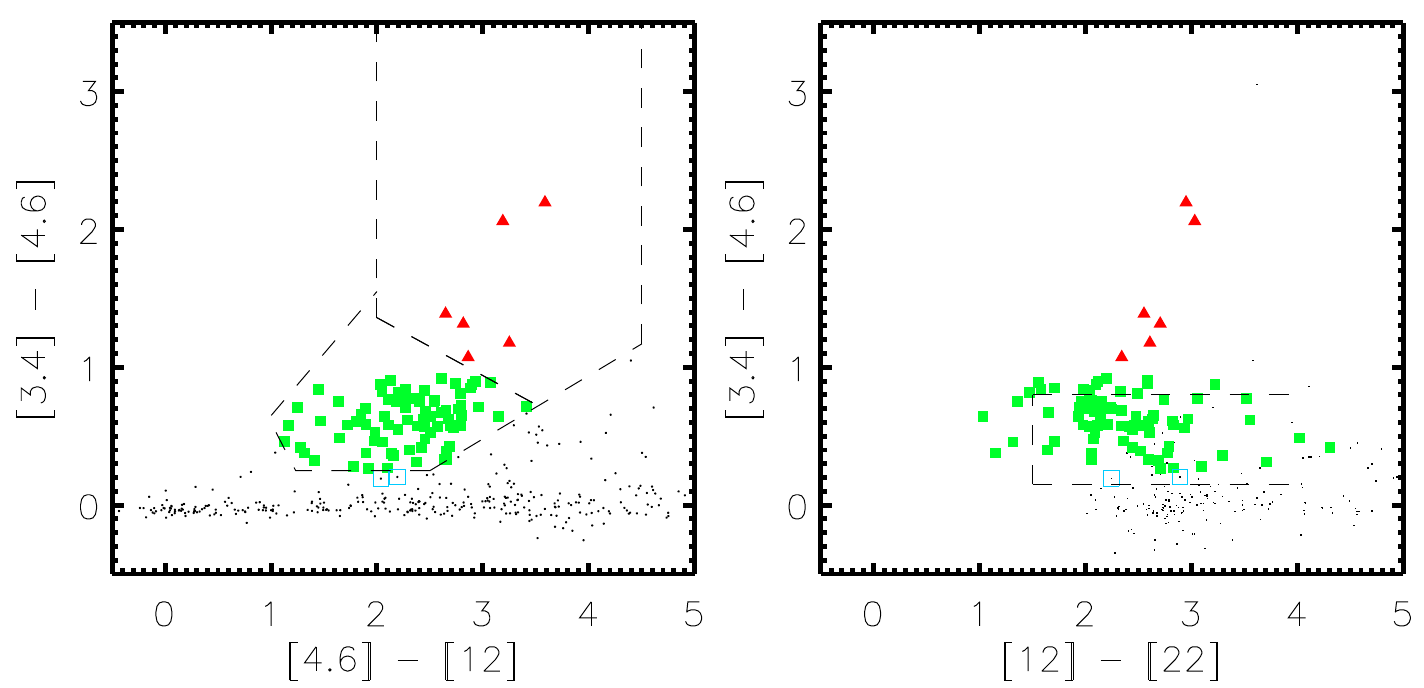}
\caption{Color-color diagrams of stars in the Rosette Nebula obtained 
from AllWISE \citep{C13}. The regions outlined by dashed lines are the 
color criteria for YSO classification \citep{KL14}. The other symbols are 
the same as Figure~\ref{afig1}.}\label{afig2}
\end{figure}

The AllWISE catalogue \citep{C13} was used to identify more YSOs 
spread over the Rosette Nebula (a $6^{\circ} \times 6^{\circ}$ region 
centered at R.A. = 06$^{\mathrm{h}}$ 31$^{\mathrm{m}}$ 
$55\fs00$, Decl. = $+04^{\circ}$ 56$^{\prime}$ $30\farcs0$, J2000). 
All YSOs and the other sources were classified according to the scheme 
of \citet{KL14}. This catalogue contains a large number of spurious sources. 
All the criteria examined by \citet{KL14} 
were applied to minimize the contamination by these spurious sources. 
The candidates of AGNs and star-forming galaxies were then sorted 
out from the color-magnitude cuts. Class I, Class II, and YSOs with a
transitional disk were identified from color-color diagrams (Figure~\ref{afig2}). 
Among these young stars, the candidates of asymptotic giant branch stars and classical 
Be stars were excluded in the list of YSO. In the end, we found six Class I, 
76 Class II, and two YSOs with a transitional disk in total. Among them, 
a total of 40 YSO candidates were found in the field of view of the 
{\it Spitzer} observation \citep{BMR07}. This number is lower than that 
identified from the {\it Spitzer} data because of the lower 
sensitivity of the WISE mission \citep{WEM10}.

A total of 338 YSO candidates were identified from the {\it Spitzer} 
and AllWISE data \citep{BMR07,C13}. There are 37 YSO 
candidates in common between the data sets. Our classification shows a good 
consistency except for two candidates. One star (R.A. $= 06^{\mathrm{h}} \ 
32^{\mathrm{m}} \ 11\fs98$, Decl. $= +05^{\circ} \ 00^{\prime} \ 
31\farcs0$, J2000) was classified as Class I in the {\it Spitzer} data 
and Class II in AllWISE data. This star was identified as an X-ray 
emitting star \citep{WTF08}, and thereby it is most likely a YSO. 
Similarly, the other star (R.A. $= 06^{\mathrm{h}} \ 32^{\mathrm{m}} \ 49\fs49$, 
$+04^{\circ} \ 43^{\prime} \ 37\farcs6$, J2000) was classified 
as Class I and Class II in the {\it Spitzer} and AllWISE data, 
respectively. Since there are a few stars around this star within a radius 
10$\arcsec$, the measured magnitude and colors could be affected 
by these neighboring sources. The intrinsic color variation may also 
be responsible for the discrepancy between the two classifications \citep{WGP18}.

%% For this sample we use BibTeX plus aasjournals.bst to generate the
%% the bibliography. The sample63.bib file was populated from ADS. To
%% get the citations to show in the compiled file do the following:
%%
%% pdflatex sample63.tex
%% bibtext sample63
%% pdflatex sample63.tex
%% pdflatex sample63.tex

%\bibliography{sample63}{}

\begin{thebibliography}{}
\bibitem[Allison et al.(2009)]{AGP09} Allison, R. J., Goodwin, S. P., Parker, R. J., et al. 2009, \apj, 700, L99
\bibitem[Andr\'e (2015)]{A15} Andr\'e P. 2015, HiA, 15, 31
\bibitem[Balog et al. (2007)]{BMR07} Balog, Z., Muzerolle, J., Rieke, G. H., et al. 2007, \apj, 660, 1532
\bibitem[Banerjee \& Kroupa (2013)]{BK13} Banerjee, S., \& Kroupa, P. 2013, \apj, 764, 29
\bibitem[Banerjee \& Kroupa (2015)]{BK15} Banerjee, S., \& Kroupa, P. 2015, \mnras, 447, 728
\bibitem[Bate (2009)]{B09} Bate, M. R. 2009, \mnras, 392, 590 
\bibitem[Binney \& Tremaine (1987)]{BT87} Binney, J., \& Tremaine, S. 1987, Galactic dynamics (Princeton, NJ: Princeton University Press), 747
\bibitem[Blaauw(1964)]{B64}Blaauw, A. 1964, ARA\&A, 2, 213
\bibitem[Bonatto \& Bica (2009)]{BB09} Bonatto, C., \& Bica, E. 2009, \mnras, 394, 2127
\bibitem[Bonnell et al. (2003)]{BBV03} Bonnell, I., A., Bate, M. R., \& Vine, S. G. 2003, \mnras, 343, 413
\bibitem[Bonnell et al. (2011)]{BSCB11} Bonnell I. A., Smith R. J., Clark P. C., \& Bate M. R., 2011, \mnras, 410, 2339
\bibitem[Brice\~{n}o et al.(2007)]{BPS07} Brice\~{n}o, C., Preibisch, T., Sherry, W. H., et al. 2007, Protostars and Planets V, 345
\bibitem[Braine et al. (2020)]{BHR20} Braine, J., Hughes, A., Rosolowsky, E., et al. 2020, \aap, 633, 17
\bibitem[Cambr\'{e}sy et al. (2013)]{CMF13} Cambr\'{e}sy, L., Marton, G., Feher, O., T\'{o}th L. V., \& Schneider, N. 2013, \aap, 557, 29
\bibitem[Cantat-Gaudin et al. (2018)]{C-G18} Cantat-Gaudin T. et al., 2018, \aap, 618, A93
\bibitem[Cantat-Gaudin et al. (2019)]{C-G19} Cantat-Gaudin T. et al., 2019, \aap, 626, A17
\bibitem[Castelli \& Kurucz(2004)]{CK04} Castelli, F., \& Kurucz, R. L. 2004, arXiv:astro-ph/0405087
\bibitem[Chen et al.(2007)]{CDZ07} Chen, L., de Grijs, R., \& Zhao, J. L. 2007, \aj, 134, 1368
\bibitem[Choi et al. (2016)]{CDC16} Choi, J., Dotter, A., Conroy, C., et al. 2016, \apj, 823, 102
\bibitem[Copetti et al. (2000)]{CMSC00} Copetti M. V. F., Mallmann J. A. H., Schmidt A. A., and Casta\~neda H. O. 2000, \aap, 357, 621
\bibitem[Corsaro et al.(2017)]{CLG17} Corsaro, E., Lee, Y.-N., Garc\'{i}a, R. A., et al. NatAs, 1, 64
\bibitem[Cutri et al. (2013)]{C13} Cutri, R. M., et al. 2013, yCat, 2328, 0
\bibitem[Dale et al. (2012)]{DEB12} Dale J. E., Ercolano B., \& Bonnell I. A., 2012, \mnras, 427, 2852
\bibitem[Dale et al. (2013)]{DEB13} Dale J. E., Ercolano B., \& Bonnell I. A., 2013, \mnras, 431, 1062
\bibitem[Dalessandro et al. (2021)]{DVT21} Dalessandro, E., Varri, A. L., Tiongco, M., et al. arXiv:astro-ph/2101.04133
\bibitem[Dickman et al.(1990)]{DHM90} Dickman, R. L., Horvath, M. A., \&  Margulis, M. 1990, \apj, 365, 586
\bibitem[Dobbs et al. (2011)]{DBP11} Dobbs, C. L., Burkert, A., \& Pringle, J. E. 2011, \mnras, 417, 1318
\bibitem[Donrey et al.(2008)]{DRPB08} Donley, J. L., Rieke, G. H., P\'erez-Gonz\'alez, P. G., \& Barro, G. 2008, \apj, 687, 111
\bibitem[Dotter (2016)]{D16}Dotter, A. 2016, \apjs, 222, 8
\bibitem[Efremov (1976)]{E76} Efremov, Y. N. 1976, SvAL, 4, 66
\bibitem[Elmegreen et al.(2000)]{EEPZ00}Elmegreen, B. G., Efremov, Y., Pudritz, R. E., \& Zinnecker, H. 2000, in Protostars and Planets IV, ed. V. Mannings, A. P. Boss, \& S. S. Russell (Tucson, AZ: Univ. Arizona Press), 179
\bibitem[Elmegreen \& Lada (1977)]{EL77} Elmegreen B. G., \& Lada C. J., 1977, \apj, 214, 725
\bibitem[Elson et al. (1987)]{EFF87}Elson, R. A. W., Fall, S. M., \& Freeman, K. C. 1987, \apj, 323, 54
\bibitem[Fukuda et al. (2002)]{FHS02} Fukuda, N.,  Hanawa, T., \&  Sugitani, K. 2002, \apj, 568, L127
\bibitem[Gavagnin et al.(2017)]{GBRT17} Gavagnin, E., Bleuler, A., Rosdahl, J., \& Teyssier, R. 2017, \mnras, 472, 4155
\bibitem[Gaia Collaboration et al.(2016)]{gaia16} Gaia Collaboration et al. 2016, \aap, 595, 1
\bibitem[Gaia Collaboration et al.(2018)]{gdr2} Gaia Collaboration et al. 2018, \aap, 616, A1
\bibitem[Gaia Collaboration et al. (2020)]{gedr3} Gaia Collaboration et al. 2020, arXiv:astro-ph/2012.01533
\bibitem[Gennaro et al. (2011)]{GBSH11} Gennaro, M., Brandner, W., Stolte, A., \& Henning, Th. 2011, \mnras, 412, 2469
\bibitem[Gouliermis (2018)]{G18} Gouliermis, D. A. 2018, \pasp, 130, 072001
\bibitem[Gutermuth et al. (2008)]{GMM08} Gutermuth, R. A.,  Myers, P. C.,  Megeath, S. T., et al. 2008, \apj, 674, 336
\bibitem[H\'{e}nault-Brunet et al. (2012)]{HGE12} H\'{e}nault-Brunet, V., Gieles, M., Evans, C. J., et al. 2012, \aap, 545, 1
\bibitem[Hensberge et al. (2000)]{HPV00} Hensberge, H., Pavlovski, K., \&  Verschueren, W. 2000, \aap, 358, 553
\bibitem[Heyer et al. (2006)]{HWB06} Heyer, M. H., Williams, J. P., \& Brunt, C. M. 2006, \apj, 643, 956
\bibitem[Hills (1980)]{H80} Hills J. G., 1980, \apj, 235, 986
\bibitem[Koenig et al.(2008)]{KAG08} Koenig X. P., Allen L. E., Gutermuth R. A., et al. 2008, \apj, 688, 1142
\bibitem[Koenig et al.(2012)]{KLB12} Koenig, X. P., Leisawitz, D. T., Benford, D. J., et al. 2012, \apj, 744, 130
\bibitem[Koenig \& Leisawitz (2014)]{KL14} Koenig, X. P., \& Leisawitz, D. T. 2014, \apj, 791, 131
\bibitem[Kounkel et al. (2018)]{KCS18} Kounkel, M., et al. 2018, \aj, 156, 84
\bibitem[Kroupa (2001)]{K01} Kroupa, P. 2001, \mnras, 322, 231
\bibitem[Kroupa et al.(2001)]{KAH01}Kroupa, P., Aarseth, S. \& Hurley, J. 2001, \mnras, 321, 699
\bibitem[Kruijssen (2012)]{K12} Kruijssen J. M. D. 2012, \mnras, 426, 3008
\bibitem[Kuhn et al. (2019)]{KHS19} Kuhn, M. A., Hillenbrand, L. A., Sills, A., Feigelson, E. D., \& Getman, K. V. 2019, \apj, 870, 32
\bibitem[Kurtz \& Mink(1998)]{KM98}Kurtz, M. J. \& Mink, D. J. 1998, \pasp, 110, 934
\bibitem[Lada (1987)]{L87} Lada, C. J. 1987, in IAU Symposium, Vol. 115, Star Forming Regions, ed. M. Peimbert \& J. Jugaku, 1
\bibitem[Lada \& Lada(2003)]{LL03} Lada, C., \& Lada, E. 2003, \araa, 41, 57
\bibitem[Lada et al.(1984)]{LMD84} Lada, C., Margulis, M., \& Dearborn, D. 1984, \apj, 285, 141
\bibitem[Lane et al. (2009)]{LKL09} Lane, R. R., Kiss, L. L., Lewis, G. F., et al. 2009, \mnras, 400, 917
\bibitem[Larson (1981)]{L81} Larson R. B. 1981, \mnras, 194, 809
\bibitem[Lee et al. (2020)]{LJL20} Lee, Y.-H., Johnstone, D., Lee, J.-E., et al. 2020, \apj, in press
\bibitem[Li (2005)]{L05} Li, J. Z. 2005, \apj, 625, 242
\bibitem[Lim et al. (2013)]{LCS13} Lim, B., Chun, M.-Y., Sung, H., et al. 2013, \aj, 145, 46
\bibitem[Lim et al. (2020)]{LHY20} Lim, B., Hong, J., Yun, H.-S., et al. 2020, \apj, 899, 121
\bibitem[Lim et al. (2019)]{LNGR19} Lim, B., Naz\'e, Y., Gosset, E., \& Rauw, G. 2019, \mnras, 490, 440
\bibitem[Lim et al. (2015)]{LSB15} Lim, B., Sung, H., Bessell, M. S., et al. 2015, \aj, 149, 127
\bibitem[Lim et al. (2018)]{LSB18} Lim, B., Sung, H., Bessell, M. S., et al. 2018, \mnras, 477, 1993
\bibitem[Lim et al. (2014)]{LSK14} Lim, B., Sung, H., Kim, J. S., Bessell, M. S., \& Park, B.-G. 2014, \mnras,
443, 454
\bibitem[Lindegren et al. (2020)]{LBB20} Lindegren, L., Bastian, U., Biermann, M., et al. 2020, arXiv:2012.01742
\bibitem[Mackey et al. (2013)]{MDFY13} Mackey, A. D., Da Costa, G. S., Ferguson, A. M. N., \& Yong, D. 2013, \apj, 762, 65
\bibitem[Ma\'iz Apell\'aniz et al. (2013)]{MSM13} Ma\'iz Apell\'aniz, J., Sota, A., Morrell, N. I.,  et al. 2013, in Massive Stars: From Alpha to Omega, 198 
\bibitem[Mahy et al. (2009)]{MNR09} Mahy, L., Naz\'e, Y., Rauw, G., et al. 2009, \aap, 502, 937
\bibitem[Mapelli (2017)]{M17} Mapelli, M. 2017, \mnras, 467, 3255
\bibitem[Marco \& Negueruela (2002)]{MN02} Marco A., \& Negueruela I. 2002, A\&A, 393, 195
\bibitem[McMillan(2017)]{Mc17} McMillan, P. J. 2017, \mnras, 465, 76
\bibitem[Mel’nik \& Dambis (2017)]{MD17} Mel’nik A. M., \& Dambis A. K., 2017, \mnras, 472, 3887
\bibitem[Mermilliod \& Paunzen (2003)]{MP03} Mermilliod, J.-C., \& Paunzen, E. 2003, \aap, 410, 511
\bibitem[Miller \& Scalo(1978)]{MS78}Miller, G. E. \& Scalo, J. M. 1978, \pasp, 90, 506
\bibitem[Mu\v{z}i\'c et al. (2019)]{MSR19} Mu\v{z}i\'c, K., Scholz, A., Ram\'irez K. P., et al. 2019, \apj, 881, 79
\bibitem[Ogura \& Ishida (1981)]{OI81} Ogura, K., \& Ishida, K. 1981, \pasj, 33, 149 
\bibitem[Padoan et al.(2001)]{PJGN01} Padoan P., Juvela M., Goodman A. A., Nordlund \AA., 2001, \apj, 553, 227
\bibitem[Panwar et al.(2019)]{PSP19} Panwar, N., Samal, M. R., Pandey, A. K., Singh, H. P., \& Sharma, S. 2019, \aj, 157, 112
\bibitem[Park \& Sung (2002)]{PS02} Park, B.-G., \& Sung, H. 2002, \aj, 123, 892
\bibitem[Parker \& Wright(2016)]{PW16} Parker, R. J., \& Wright, N. J. 2016, \mnras, 457, 3430
\bibitem[P\'erez et al. (1987)]{PTW87} P\'erez, M. R., The, P. S., \& Westerlund, B. E. 1987, \pasp, 99, 1050
\bibitem[P\'erez (1991)]{P91} P\'erez, M. R. 1991, \rmxaa, 22, 99
\bibitem[Phelps \& Lada (1997)]{PL97} Phelps, R. L., \& Lada, E. A. 1997, \apj, 477, 176
\bibitem[Pineda et al.(2008)]{PCG08} Pineda, J. E., Caselli, P., \& Goodman, A. A. 2008, \apj, 679, 481
%\bibitem[Platais et al. (2020)]{PRB20} Platais, I., Robberto, M., Bellini, A., et al. 2020, \aj, 159, 272
\bibitem[Plummer (1911)]{P11} Plummer, H. C. 1911, \mnras, 71, 460
\bibitem[Portegies Zwart, McMillan \& Gieles(2010)]{PMG10} Portegies Zwart S. F., McMillan S. L. W., \& Gieles M., 2010, \araa, 48, 431
\bibitem[Poulton et al. (2008)]{PRG08} Poulton, C. J., Robitaille, T. P., Greaves, J. S., et al. 2008, \mnras, 384, 1249
\bibitem[Porras et al.(2003)]{PCA03} Porras, A., Christopher, M., Allen, L., et al. 2003, \aj, 126, 1916
\bibitem[Reed (2003)]{R03} Reed, B. C. 2003, \aj, 125, 2531
\bibitem[Rom\'an-Z\'u\~{n}iga et al. (2008a)]{REFL08} Rom\'an-Z\'u\~{n}iga. C. G., Elston, R., Ferreira, B., \& Lada, E. A. 2008, \apj, 672, 861
\bibitem[Rom\'an-Z\'u\~{n}iga \& Lada (2008b)]{RL08} Rom\'an-Z\'u\~{n}iga, C. G., \& Lada, E. A. 2008, in Handbook of Star Forming Regions, Volume I: The Northern Sky ASP Monograph Publications, Vol. 4, ed. B. Reipurth (San Francisco, CA: ASP), 928
\bibitem[Rom\'an-Z\'u\~{n}iga et al. (2019)]{R-Z19} Rom\'an-Z\'u\~{n}iga, C. G., Roman-Lopes, A., Tapia, M., Hern\'andez, J., Ram\'irez-Preciado V. 2019, \apj, 87, L12
\bibitem[Rosolowsky et al. (2003)]{REPB03} Rosolowsky, E., Engargiola, G., Plambeck, R., \& Blitz, L. 2003, \apj, 599, 258
\bibitem[Sabbi et al. (2012)]{SLG12} Sabbi, E., Lennon, D. J., Gieles, M., et al. 2012, \apj, 754, L37
\bibitem[Salpeter (1955)]{S55} Salpeter, E. E. 1955, \apj, 121, 161
\bibitem[Schnee et al. (2007)]{SCG07} Schnee, S., Caselli, P., Goodman, A., et al. 2007, \apj, 671, 1839
\bibitem[Schneinder et al. (2010)]{SMB10} Schneinder, N., et al. 2010, \aap, 518, L83
\bibitem[Schoettler et al. (2019)]{SPA19} Schoettler, C., Parker, R. J., Arnold, B., et al. 2019, \mnras, 487, 4615
\bibitem[Scoville et al. (1986)]{SSS86} Scoville, N. Z., Sargent, A. I., Sanders, D. B., et al. 1986, \apj, 303, 416
\bibitem[Sharma et al. (2007)]{SPO07} Sharma S., Pandey A. K., Ojha D. K., et al. 2007, \mnras, 380, 114
\bibitem[Sicilia-Aguilar et al. (2004)]{SHB04} Sicilia-Aguilar, A., Hartmann, L. W., Brice\~{n}o, C., Muzerolle, J., \& Calvet, N. 2004, \aj, 128, 805
\bibitem[Skiff (2009)]{S09} Skiff B. A., 2009, Catalogue of Stellar Spectral Classifications, VizieR Catalog B/mk/mktypes (Lowell Observatory), 1, 2023
\bibitem[Smith et al. (2006)]{SHMP06} Smith, H. A., Hora, J. L., Marengo, M., \& Pipher, J. L. 2006, \apj, 645, 1264
\bibitem[Sneden(1973)]{S73}Sneden, C. A. 1973, Ph.D. Thesis, The University of Texas at Austin, USA
\bibitem[Sung et al. (2008)]{SBC08} Sung, H., Bessell, M. S., Chun, M.-Y., Karimov, R., \& Ibrahimov, M. 2008, \aj, 135, 441
\bibitem[Sung et al.(2017)]{SBC17} Sung, H., Bessell, M. S., Chun, M.-Y., et al. 2017, \apjs, 230, 3
\bibitem[Sung et al. (2009)]{SSB09} Sung, H., Stauffer, J. R., \& Bessell, M. S. 2009, \aj, 138, 1116
\bibitem[Szentgyorgyi et al.(2011)]{SFC11}Szentgyorgyi, A., Furesz, G., Cheimets, P., et al. 2011, \pasp, 123, 1188
\bibitem[Tasker (2011)]{T11} Tasker, E. J. 2011, \apj, 730, 11
\bibitem[Tonry \& Davis(1979)]{TD79}Tonry, J. \& Davis, M. 1979, \aj, 84, 1511
\bibitem[Tutukov (1978)]{T78} Tutukov A. V.,1978, \aap, 70, 57
\bibitem[Wang \& Chen (2019)]{WC19} Wang, S., \& Chen, X. 2019, \apj, 877, 116
\bibitem[Wang et al. (2009)]{WFT09} Wang, J., Feigelson, E. D., Townsley, L. K., et al. 2009, \apj, 696, 47
\bibitem[Wang et al. (2010)]{WFT10} Wang, J., Feigelson, E. D., Townsley, L. K., et al. 2010, \apj, 716, 474
\bibitem[Wang et al. (2008)]{WTF08} Wang, J., Townsley, L. K., Feigelson, E. D., et al. 2008, \apj, 675, 464
\bibitem[Ward et al. (2020)]{WKR20} Ward, J. L., Kruijssen, J. M. D., \& Rix, H. W. 2020, \mnras, 495, 663
\bibitem[Wolk et al.(2018)]{WGP18} Wolk, S. J., G\"unther, H. M., Poppenhaeger, K., et al. 2018, \aj, 155, 9
\bibitem[Wright et al. (2010)]{WEM10}Wright E. L. et al., 2010, \aj, 140, 1868 
\bibitem[Ybarra et al. (2013)]{YLR13} Ybarra, J. E., Lada, E. A., Rom\'an-Z\'u\~{n}iga, C. G., et al. 2013, \apj, 769, 140
\end{thebibliography}
%\bibliographystyle{aasjournal}

%% This command is needed to show the entire author+affiliation list when
%% the collaboration and author truncation commands are used.  It has to
%% go at the end of the manuscript.
%\allauthors

%% Include this line if you are using the \added, \replaced, \deleted
%% commands to see a summary list of all changes at the end of the article.
%\listofchanges

\end{document}